\documentclass[twocolumn,preprint]{aastex63}

\usepackage{amsmath,amstext}
\usepackage[T1]{fontenc}
\usepackage{apjfonts} 
\usepackage{natbib}
\usepackage{microtype}
\usepackage{longtable}
\usepackage{verbatim}
\usepackage{hyperref}
\hypersetup{
    colorlinks=true,
    linkcolor=blue,
    filecolor=magenta,      
    urlcolor=cyan,
}

\newcommand{\hst}{\textit{HST}}
\newcommand{\HST}{\textit{HST}}

\newcommand{\lsim}{\lesssim}
\newcommand{\gsim}{\gtrsim}

\newcommand{\spitzer}{\textit{Spitzer}}
\newcommand{\jwst}{\textit{JWST}}
\newcommand{\threedhst}{\hbox{3D-HST}}

\newcommand{\zfif}{\hbox{$z_{50}$}}

\newcommand{\Lsig}{\hbox{$\log(\Sigma_1)$}}

\newcommand{\Mkpcsq}{\hbox{$M_\odot\ \mathrm{kpc}^{-2}$}}

\newcommand{\editone}[1]{\textcolor{black}{#1}}
\newcommand{\edittwo}[1]{\textcolor{black}{#1}}

\begin{document}
\definecolor{aggiemaroon}{HTML}{500000}

\title{\large \bf CLEAR II: Evidence for Early Formation of the Most Compact Quiescent Galaxies at High Redshift}

\author[0000-0001-8489-2349]{Vicente Estrada-Carpenter}
\affiliation{Department of Physics and Astronomy, Texas A\&M University, College
Station, TX, 77843-4242 USA}
\affiliation{George P.\ and Cynthia Woods Mitchell Institute for
 Fundamental Physics and Astronomy, Texas A\&M University, College
 Station, TX, 77843-4242 USA}
 
\author[0000-0001-7503-8482]{Casey Papovich}
\affiliation{Department of Physics and Astronomy, Texas A\&M University, College
Station, TX, 77843-4242 USA}
\affiliation{George P.\ and Cynthia Woods Mitchell Institute for
 Fundamental Physics and Astronomy, Texas A\&M University, College
 Station, TX, 77843-4242 USA}

\author[0000-0003-1665-2073]{Ivelina Momcheva}
\affil{Space Telescope Science Institute, 3700 San Martin Drive,
  Baltimore, MD, 21218 USA}

\author[0000-0003-2680-005X]{Gabriel Brammer}
\affil{Cosmic Dawn Centre, University of Copenhagen, Blegdamsvej 17, 2100 Copenhagen, Denmark}

\author[0000-0002-6386-7299]{Raymond Simons}
\affil{Space Telescope Science Institute, 3700 San Martin Drive,
  Baltimore, MD, 21218 USA}

\author[0000-0002-8584-1903]{Joanna Bridge}
\affil{Department of Physics and Astronomy, 102 Natural Science Building, University of Louisville, Louisville, KY, 40292 USA}

\author[0000-0001-7151-009X]{Nikko J. Cleri}
\affil{Department of Physics, University of Connecticut, Storrs, CT 06269, USA}

\author{Henry Ferguson}
\affil{Space Telescope Science Institute, 3700 San Martin Drive,
  Baltimore, MD, 21218 USA}

\author[0000-0001-8519-1130]{Steven L. Finkelstein}
\affil{Department of Astronomy, The University of Texas at Austin, Austin,
TX 78712, USA}

\author{Mauro Giavalisco}
\affil{Astronomy Department, University of Massachusetts,
Amherst, MA, 01003 USA}

\author[0000-0003-1187-4240]{Intae Jung}
\affil{Department of Physics, The Catholic University of America, Washington, DC, 20064 USA}
\affil{Astrophysics Science Division, Goddard Space Flight Center, Greenbelt, MD, 20771 USA}

\author[0000-0002-7547-3385]{Jasleen Matharu}
\affiliation{Department of Physics and Astronomy, Texas A\&M University, College
Station, TX, 77843-4242 USA}
\affiliation{George P.\ and Cynthia Woods Mitchell Institute for
 Fundamental Physics and Astronomy, Texas A\&M University, College
 Station, TX, 77843-4242 USA}

\author[0000-0002-1410-0470]{Jonathan R. Trump}
\affil{Department of Physics, University of Connecticut, Storrs, CT 06269, USA}

\author[0000-0001-6065-7483]{Benjamin Weiner}
\affil{MMT/Steward Observatory, 933 N. Cherry St., University of Arizona, Tucson,
AZ 85721, USA}
%
%
%
%


\begin{abstract}
The origin of the correlations between mass, morphology, quenched
fraction, and formation history in galaxies is difficult to define,
primarily due to the uncertainties in galaxy star-formation histories.
Star-formation histories are better constrained for
higher redshift galaxies, observed closer to their formation and
quenching  epochs. Here we use ``non-parametric'' star-formation
histories and a nested sampling method to derive constraints on the
formation and quenching timescales of quiescent galaxies at
$0.7<z<2.5$.  We model deep \hst\ grism spectroscopy and photometry
from the CLEAR (CANDELS Lyman$-\alpha$ Emission at Reionization)
survey. The galaxy formation redshifts, $z_{50}$ (defined as the point
where they had formed 50\% of their stellar mass) range from
$z_{50}\sim 2$ (shortly prior to the observed epoch) up to $z_{50}
\simeq 5-8$.  \editone{We find that early formation redshifts are correlated with high stellar-mass surface densities, $\log \Sigma_1 / (\Mkpcsq) >$10.25,
where  $\Sigma_1$ is the stellar mass within 1~pkpc (proper kpc).  
Quiescent galaxies with the highest stellar-mass surface density,
$\log\Sigma_1 / (M_\odot\ \mathrm{kpc}^{-2}) > 10.25$, } show a \textit{minimum} formation redshift: all such objects in our
sample have $z_{50} > 2.9$.  \editone{ Quiescent galaxies with lower surface
density, $\log \Sigma_1 / (M_\odot\ \mathrm{kpc}^{-2}) = 9.5 - 10.25$,
show a range of formation epochs ($z_{50} \simeq 1.5 - 8$), implying
these  galaxies experienced a range of formation and assembly
histories.  We argue that the surface density threshold
$\log\Sigma_1/(M_\odot\ \mathrm{kpc}^{-2})>10.25$ uniquely identifies}
galaxies that formed in the first few Gyr after the Big Bang, and we
discuss the implications this has for galaxy formation models.

%
\end{abstract}

\section{Introduction}
One of the major outstanding questions in galaxy evolution is  ``how
do massive quiescent galaxies form?''.  These galaxies exhibit many
extreme traits:  compact morphologies \citep[e.g.][]{whit12,
vanw14}, indications of rapid formation histories (including
[$\alpha$/Fe] enhancement and high star-formation rates [SFRs] at
early times) \citep[e.g.,][]{papo06,lono15,krie15,krie19}, old stellar
populations \citep{gall05,thom05,gall14,estr19}, and high overall
metallicity ($Z \simeq Z_\odot$) \citep{estr19,ferr19,krie19}.
%
%
%
%
%

Multiple theories have been proposed to explain the
inability of massive galaxies to continue star-formation.  A very 
important difference in these models is the timescale of quenching
\citep[e.g.,][]{man18}.   Studies in this area have led to the
identification of ``fast'' and ``slow''
evolutionary paths, which describe the relative rate of quenching \citep{barr13, well15, beli19}.

The  slow path applies to galaxies
that quench their star-formation by experiencing a 
gradual slowdown in
their gas accretion rates combined with the consumption or heating of their
existing gas \editone{\citep[as may be the case in the Milky Way,][]{papo15}.}
These galaxies can have compact morphologies if they  formed in the early
universe \citep[when densities were higher, e.g.][]{well15} or if they undergo
(secular) compaction events or dissipative mergers \citep{deke09,well15,barr17}.
%
%
The fast path normally requires a fast-acting compaction event (i.e.,
major mergers, extreme disk instabilities).  This can drive extreme
star-formation and/or supermassive black hole
accretion, the feedback from  which quenches star-formation.  Due to
higher gas fractions, the fast quenching path may be more common in
the early universe \citep{deke09, dadd10}.

The key difference in the physical processes of quenching is the speed
at which it occurs.  This can be studied using constraints on the galaxies'
star-formation histories (SFHs). Quenching can be correlated with
galaxy morphology if the quenching mechanism involves reorganization
of the galaxies' stellar component (such as compaction), or a
natural consequence of ``inside-out'' growth combined with disk fading
e.g., \citealt{lilly16}). Therefore, deriving the SFHs and comparing
them to the morphologies of galaxies has the potential to constrain
the quenching mechanisms.

Here, we aim to constrain the SFHs of a large sample of massive
quiescent galaxies ($\log M_\ast / M_\odot > 10.5$) at $z \gtrsim 1$
and study these  as a function of morphology.
Throughout we use a cosmology with $\Omega_{m,0}=0.3$,
$\Omega_{\Lambda,0}=0.7$, and $H_0 = 70$~km s$^{-1}$ Mpc$^{-1}$. 

\section{Data \label{Sec_data}}

\begin{figure*}[th]
\epsscale{1.15}
\plotone{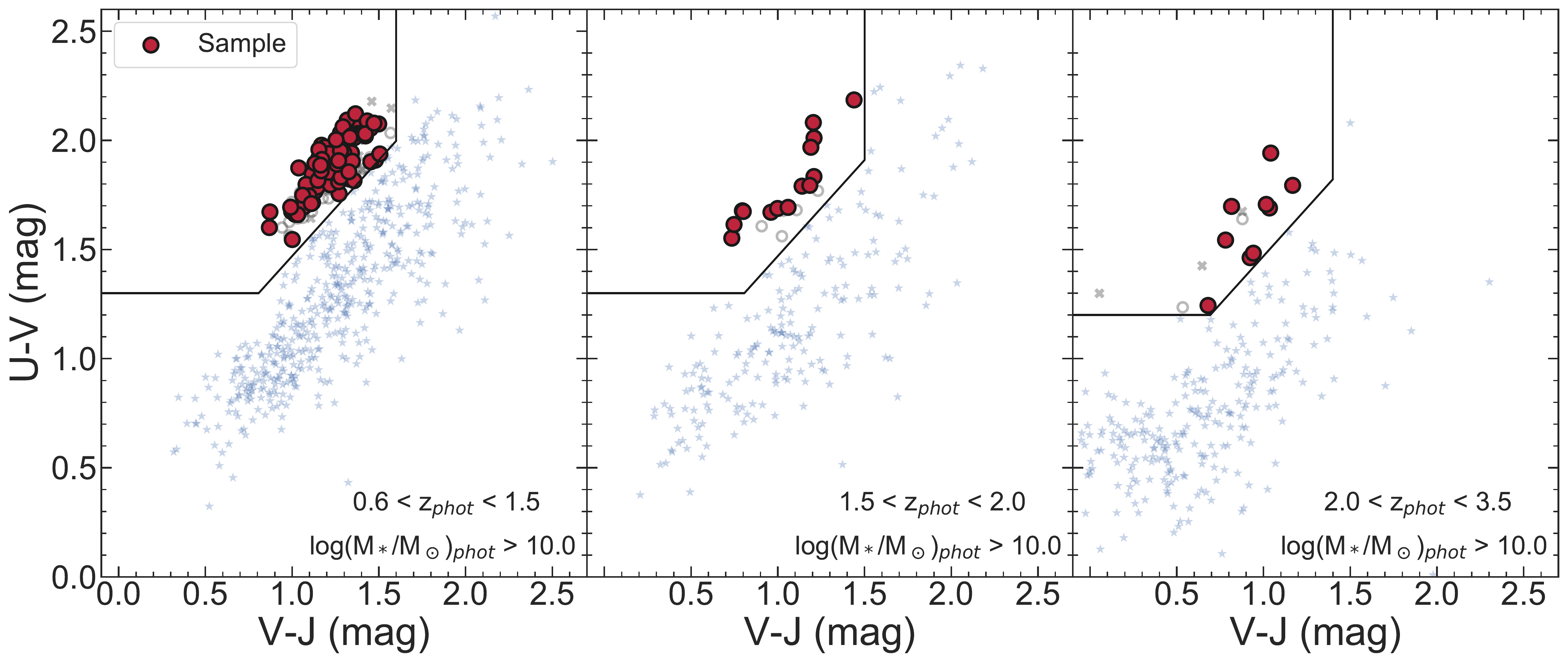}
\caption{\editone{$V-J$ versus $U-V$ rest-frame color-color diagram
    (``$UVJ$'' plot) of all CLEAR galaxies with 0.6 $<$ $z_\mathrm{phot}$ $<$ 3.5
and $\log (M_\mathrm{phot}  / M_{\odot})  > 10.0$. Galaxies which fall into the 
quiescent wedge (upper left region in each panel) are candidate
quiescent galaxies and constitute our parent sample.  The red larger
points show galaxies that satisfy our final sample selection of 0.7 $<$ $z_\mathrm{grism}$ $<$ 2.5
and $\log (M_\mathrm{grism} / M_{\odot})   > 10.5$ (and satisfy our
X-ray and 24~$\mu$m selection, see Section
\ref{Sec_samp}).  Blue stars show galaxies that fail the quiescent-galaxy selection
(i.e., they are star-forming galaxies). Grey X's mark quiescent
galaxies that were rejected \edittwo{(mostly because they have
grism-redshits outside our final redshift range).}   Open grey circles
show quiescent galaxies that are rejected for falling under our final stellar-mass limit 
$\log(M_*/M_\odot) > 10.5$.}
\label{fig_uvj}}
\end{figure*} 

We use data from the CLEAR survey (a Cycle 23 \hst\ program, PI:
Papovich), which consists of deep (12 orbit) \HST/WFC3 G102 slitless
grism spectroscopy covering $0.8-1.2$~\micron\ within 12 fields split
between the CANDELS GOODS-North  (GN) and GOODS-South (GS) fields
\citep[see,][]{grog11,koek11,estr19}. These fields overlap with the
\threedhst\ GOODS fields \citep{momc16}, which provide shallow (2
orbit depth) slitless G141 grism coverage from $1.1-1.65$~\micron.
The galaxies of interest lie at $0.7 < z_{grism} < 2.5$, where our
spectral  coverage includes many metallicity and age features.  These
include the 4000 \AA\ break, Balmer lines (H$\delta$, H$\gamma$,
H$\beta$, H$\alpha$), Ca HK, Mg$b$, and other absorption features.
We also utilize the broadband photometry available (using an updated
catalog from \cite{skel14} that includes photometry in the $Y$-band
from \hst/WFC3 in F098M or F105W, see \citealt{estr19}  and CLEAR
collaboration, in prep).  The broadband photometric data spans
0.3--8~\micron\ (rest-frame UV to near-IR)  allowing for better SFH
constraints. \editone{ We include all bands available in
\edittwo{\cite{skel14}} (now including the WFC3 F098M and F105W data),
with the exceptions of the MOIRCS $J$, and Suprime-cam $I$, $Z$-bands
in GOODS-N, and with the exceptions of the ISAAC $JHK$, the ESO/WFI
$I$-band, and the IA768, IA797 filters in GOODS-S.  These bands
consistently showed large biases in the flux calibration (up to
$0.3$~mag) compared to residuals between the galaxies' data and our
best-fit models.  While these flux-calibration offsets were consistent
with those reported by \citeauthor{skel14}, we found very larger
scatter, which made their flux calibration uncertain.   In all cases
the excluded bands are significantly shallower than other bands that
cover these wavelengths (by up to $1-1.5$~mag), and excluding these
bands had no impact on our final fits.} 

\begin{figure*}[th]
\epsscale{1.15}
\plotone{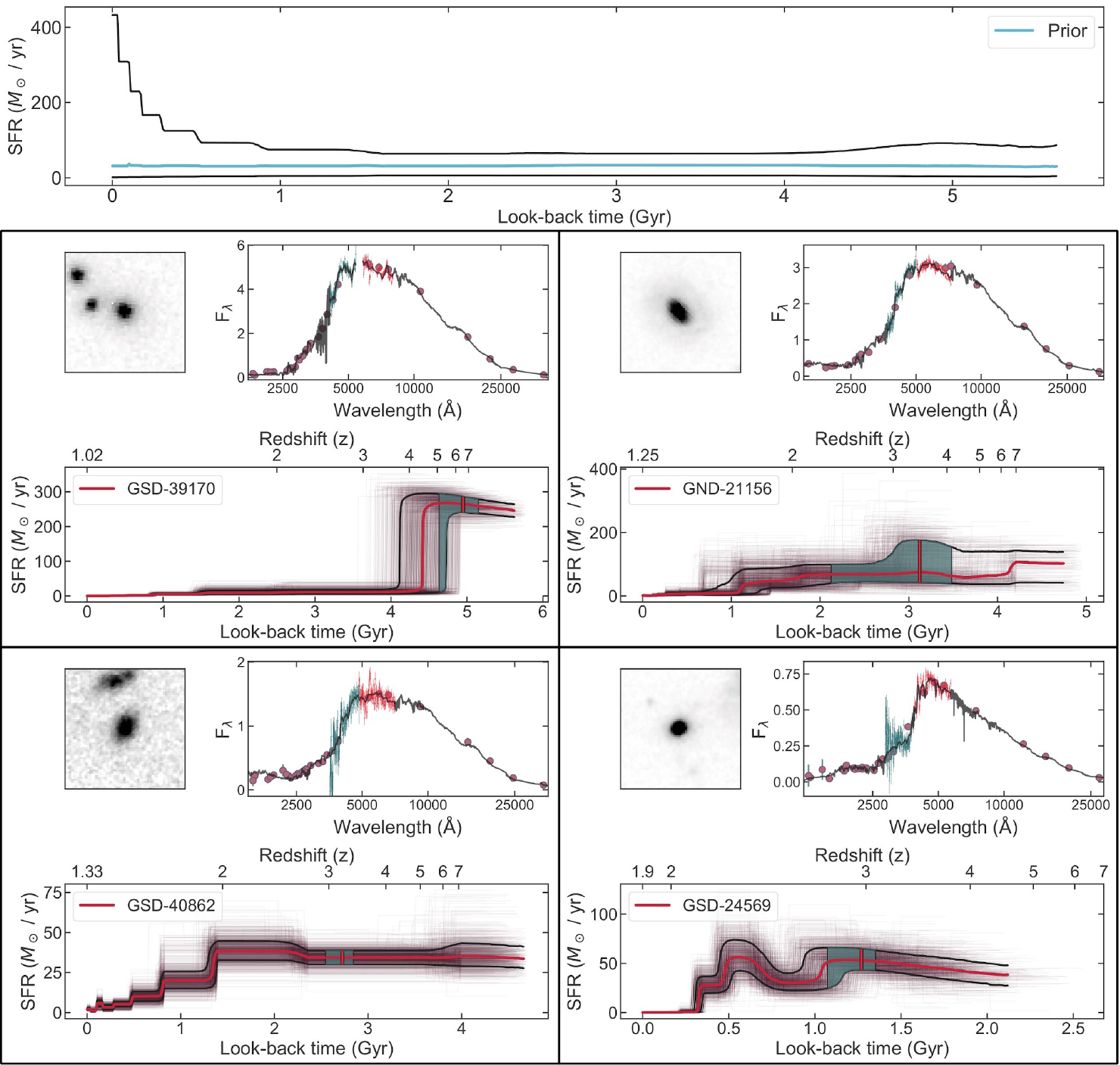}
\caption{Example spectral energy distribution (SED) fits to galaxies
from our sample. Each set of bottom four sub-panels shows results for
one galaxy (with CLEAR IDs labeled). \editone{The top sub-panel shows
the shape of the prior used for the SFH (median in blue and the 68\%
credible region in black). The prior shown is specifically for a
galaxy at $z=1.02$ with stellar mass $\log M/M_\odot = 11.40$ (like
GSD-39170), and changes to the redshift and stellar mass affect the
span of the star-formation history (set by redshift) and SFR
normalization (set by mass); the overall shape of the prior is the
same for all galaxies. In each of the following sub-panels, the
top-left sub-panels shows a 4" $\times$ 4" F160W image centered on the
galaxy}. The top right sub-panels show the full SED including the
broadband photometry (purple circles) and WFC3 grism spectra (blue
line: WFC2/G102;  red line:  WFC3/G141) along with median FSPS stellar
population model from the posterior (black line).  The bottom figure
in each sub-panel shows the derived star-formation history (SFH).  The
purple lines show individual draws for the SFH, the thick red line
shows the median, and the thick black lines show the 68\% credible
interval. The vertical red line shows \zfif, the formation redshift
(where 50\% of the stellar mass had formed), and the green-shaded
region shows the 68\% highest density region on \zfif. \edittwo{In
  Appendix~\ref{sec_appendix_online} we provide a
  hyperlink to, and a description of, an online appendix that contains similar fits and
  information for all the galaxies in our sample.} 
\label{fig_data}}
\end{figure*}

Figure~\ref{fig_data} shows examples of the full data coverage for the
broadband photometry and grism spectroscopy.  We used the
\textit{Grism redshift \& line} analysis software
Grizli\footnote{https://github.com/gbrammer/grizli} for spectral
extractions and grism forward modeling.  For our analysis, we include
all \hst/WFC3 grism data available. The primary dataset is from CLEAR,
but we include additional WFC3/G102 data from GO 13420 (PI: Barro) and
Faint Infrared Grism Survey (FIGS) \citep{pirz17} when these overlap
with galaxies in the CLEAR fields, as these additional data help to
reduce contamination and increase the overall signal to noise of the
grism data.  \editone{Due to the nature of grism data there are
instances when the spectra of our galaxies show residual contamination
from the spectra of nearby objects (especially in the case that the
nearby objects are significantly brighter in flux).  For our sample,
we visually inspected the individual beams of each object.  In cases
where we observed any residual contamination, we either removed those
beams or masked the residual emission.  This affected individual beams
in 9\% of the objects in our sample.}  \edittwo{The residuals from
contamination subtraction are frequently worse in the G141 spectra.
This is primarily due to the fact that these data are taken with only
a single \hst\ ORIENT.  Therefore the contamination (collisions from
the spectra of nearby galaxies) is modeled in only a single role angle, where
multiple role angles improve the correction by modeling the galaxy
spectra at independent locations \citep[see discussion in,
e.g.,][]{estr19}.  In part, this was one reason that we included in our
models additional (nuisance) parameters that allow for a bias or
tilt to the grism data (see Section~\ref{section_modeling}).  The
residuals from contaminating spectra are less severe in the G102
spectra, which include multiple orbits (at least 3 ORIENTs).   As
discussed below, we include with this \textit{Paper} an online,
interactive appendix that shows the data for each of the galaxies in
our sample, see also Appendix~\ref{sec_appendix_online}.  }

\editone{For our analysis below, we also make use of MIPS 24 $\micron$ data
  for the GOODS-N and GOODS-S fields from the GOODS Spitzer Legacy program
  \citep[PI: M. Dickinson, see,][]{magn11}.   We use here an updated
  catalog derived from photometry derived using prior source positions
  from \spitzer/IRAC (using the same procedures and methods identical
  to that of \edittwo{\cite{magn11}};
  [H.~Inami and M.~Dickinson, private communication]).  These catalogs
  are also discussed in \citet{papo15}.  }
%
%

\editone{\subsection{Sample Selection \label{Sec_samp}}}


Following \cite{estr19}, we select quiescent galaxies using a rest-frame
$(U-V)$--$(V-J)$ color-color diagram ($UVJ$) selection
\citep[see][]{whit11},
\begin{eqnarray}
(U-V)\ &\ge\ & 0.88 \times (V-J) + 0.59,\nonumber\\
(U-V)\ \ge\ 1.3,\ (V-J)\ &\leq\ & 1.6\ [0.0 < z < 1.5],\nonumber\\
(U-V)\ \ge\ 1.3,\ (V-J)\ &\leq\ & 1.5\ [1.5 < z < 2.0],\\
(U-V)\ \ge\ 1.2,\ (V-J)\ &\leq\ & 1.4\ [2.0 < z < 3.5]\nonumber
\end{eqnarray} 
as seen  in Figure \ref{fig_uvj}.
For our parent sample we select galaxies with 0.6 $<$ $z_\mathrm{phot} $
$<$ 3.5 and $\log (M_\mathrm{phot}/M_{\odot}) > 10.0$ using updated
photometric redshifts ($z_\mathrm{phot}$), stellar masses
($M_\mathrm{phot}$), and rest-frame colors derived from the broad-band
photometry derived from
EAZY-py\footnote{https://github.com/gbrammer/eazy-py}.  Our parent sample \edittwo{then consists of 174}
candidate quiescent galaxies using these selection criteria.  These
are shown in Figure~\ref{fig_uvj}, subdivided by photometric
redshift. We use the redshifts from the broad-band data to select the
parent sample even when we have redshifts from grism data using the
Grizli extractions ($z_{\mathrm{grizli}}$).  
This is because it is possible for 
Grizli to misidentify emissions lines, which occur either in low
signal-to-noise data, or in cases where objects have residual 
contamination (e.g., emission lines from nearby objects which are
removed post extraction as explained in Section \ref{Sec_data}).

\edittwo{ We therefore use the EAZY-py fits to the broad-band data to 
define our initial sample (from $z_\mathrm{phot}$ and the rest-frame
$U-V$ and $V-J$ colors) and then subsequently refine our sample using the 
fits to the broad-band photometry and both the G102 and G141 grism data from our 
analysis below ($z_\mathrm{grism}$, see Section~\ref{sec_fit} below).
Here, we provide some comparisons between the different redshifts.  
Comparing our adopted redshifts ($z_{\mathrm{grism}}$) to either those
from Grizli ($z_{\mathrm{grizli}}$) or to those from the broad-band
photometry alone ($z_{\mathrm{phot}}$), the difference is small.  We
find a small scatter for the redshifts derived from the grism data, with
$\sigma(z_{\mathrm{grizli}}-z_{\mathrm{grism}}) \sim 0.008$. }
\edittwo{Fewer than 8\% of the galaxies (13 of 174) show differences
  in redshift as large as $|z_{\mathrm{grizli}}-z_{\mathrm{grism}}| > 0.2$,
  and this appears to be the result of the misidentification 
  of weak emission lines where the grism data is noisy. The difference between our adopted
  redshifts ($z_{\mathrm{grism}}$) and those from the broad-band
  photometry alone ($z_{\mathrm{phot}}$) have larger scatter,
  $\sigma(z_{\mathrm{phot}}-z_{\mathrm{grism}}) \sim 0.03$, but this
  is consistent with the uncertainty of the photometric redshifts derived
  from broad-band photometry compared to spectroscopy \citep[see,  e.g.,][]{dahlen13,skel14,stra16}. }

\editone{ We then apply a secondary sample selection using the results
from our new stellar population fits to the broad-band data and grism
data (see Section~\ref{sec_fit} below).  First we remove 12 galaxies
that had poor quality grism spectra, either because they had low SNR (
< 1 pixel$^{-1}$), had severe contamination from nearby objects,
and/or fell  near the edge of the WFC3 grism field (where they had
$<$30\% spectroscopic coverage in the grism data).  We then refine the
selection to include only galaxies with 0.7 $<$ $z_{\mathrm{grism}}$
$<$ 2.5 and $\log (M_\mathrm{grism}/M_{\odot}) > 10.5$, where the
\texttt{grism} subscript denotes quantities deriving using our fits to
the broad-band photometry and G102+G141 grism data (see
Section~\ref{sec_fit}).   The redshift range is used to ensure that
the \hst/WFC3 G102+G141 data include important rest-optical spectral
features that are sensitive to age and metallicity (see below), while
the stellar mass limit corresponds to (approximately) a volume limited
sample limited in stellar mass $\log (M_{\mathrm{grism}}/M_{\odot}) >
10.5$ over this redshift range for our SNR requirement. Furthermore,
the bias and scatter between the stellar masses from EAZY-py (used for
the parent sample) and the grism-derived method (used for the final
sample) are small ($0.07$~dex and $0.05$~dex, respectively) so this
does not affect our final sample which uses a higher stellar-mass
limit. We then removed X-ray sources by cross-matching our catalog
with sources with $r \leq 0.5$\arcsec\ within any source in the the 2
Ms Chandra Deep Field-North Survey \citep{xue16} and 7 Ms Chandra Deep
Field-South Survey catalogs \citep{luo17}.  We also incorporate
morphological information using results from Sersic-fits, derived
using GALFIT \citep{peng02}, from \citet[]{vanw14}.  We  remove
galaxies with a fit quality flag of 3 (or ``no fit''). Finally we
limit our sample to a stellar mass surface density ($\Sigma_1$) of
\Lsig\ $>$ 9.6 to remove potential satellites. The final sample
passing all  our selection criteria includes 98 quiescent galaxies.
We show these as large red symbols in Figure \ref{fig_uvj}. }


\editone{Several previous studies \citep[e.g.,][]{whit13, fuma14} have
shown that the $UVJ$ selection of quiescent galaxies is susceptible to
contamination from dust-reddened star-forming galaxies. We tested for
this possibility in our own sample by cross-correlating the sources in
our catalog against those in the MIPS 24~$\micron$ data for these
fields.  Of our 98 quiescent galaxies we find matches for 15 of our
galaxies within 0.5\arcsec.  Because the MIPS 24~$\mu$m point-spread
function (FWHM $\simeq$6\arcsec) is substantially  larger than that of
\hst/WFC3 (FWHM $\simeq 0\farcs2$) we inspected the sources visually
using the \hst/WFC3 images (F125W, F160W bands), \spitzer/IRAC images
(3.6, 4.5, 5.8, 8.0~$\mu$m bands)  and MIPS 24~$\mu$m image.  From
this, we determined that 9/15 of the 24~$\mu$m detections are likely a
result of flux from nearby sources (as evidenced from the fact that
the nearby neighbor is brighter in the IRAC data).  We therefore do
not remove these galaxies from our sample.  In the remaining 6/15 of
the 24~$\mu$m sources, only two have SNR(24$\mu$m) $>$ 5.   For
completeness, we keep these galaxies in our sample, however, we find
that excluding them has no impact on our conclusions as they span a
range of stellar mass surface density and formation redshift see
below).   In addition, all our  galaxies have derived specific SFRs
(sSFR; averaged over the last 100 Myr) from the broad-band photometry
and grism data of \edittwo{$\log (\mathrm{sSFR} / \mathrm{yr^{-1}}) 
\lesssim-10.2$,} consistent with them being 
quiescent as they all lie at least
1.5 dex below the star-forming main sequence \citep{sant17}.
Therefore, even if these objects have obscured star-formation or AGN,
it is not a significant contributor to the light dominating the \hst\
grism data and photometry, which instead appears to originate from
passively evolving stellar populations.}

\section{Methods \label{sec_fit}}
\begin{figure*}[t]
\epsscale{1.1}
\plotone{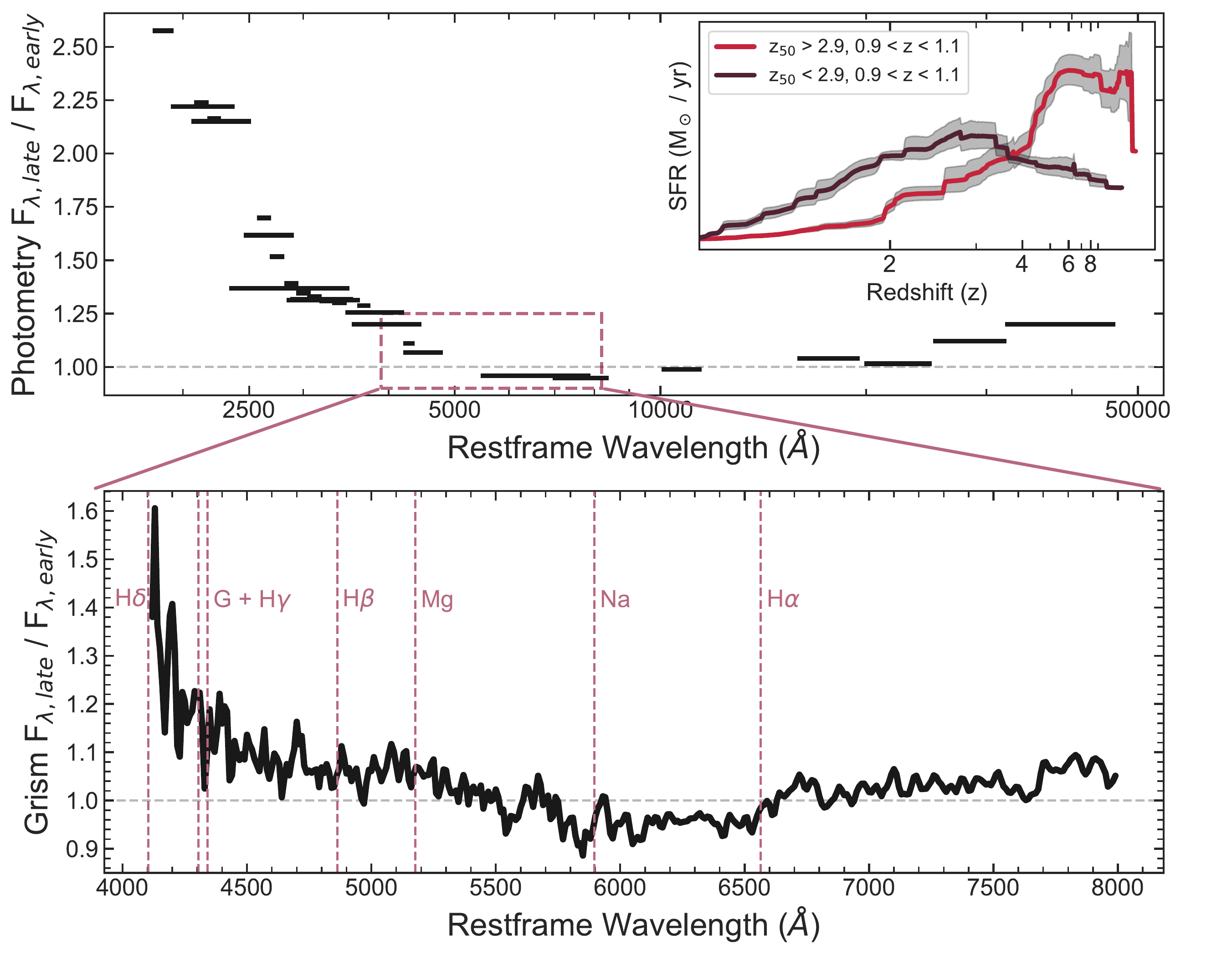}
\caption{Comparison between the photometric and spectroscopic (grism)
data for the subset of our quiescent galaxies at $0.9 < z_{grism} <
1.1$,  split by their measured formation redshift ($z_{50}$, where
50\% of their stellar mass had formed). The two groups are $z_{50} <
2.9$ (``late''  forming galaxies) and $z_{50} > 2.9$ (``early''
forming galaxies). The top plot shows the ratio of the median flux
densities measured in each broadband photometric band for the ``late''
forming sample to the  ``early'' forming sample. The biggest
difference occurs at rest UV wavelengths, which indicates the ``late''
forming galaxies show evidence of more recently formed stars (which
contributes to the lower $z_{50}$). The bottom panel shows a ratio of
their stacked combined G102 + G141 grism spectra.  \editone{Dashed
vertical lines show wavelengths of common spectral features. } For
both the top and bottom panels we normalize the stacks/medians at 6000
- 6500 \AA\ in the rest-frame. The inset in the top panel shows a mean
stack of the SFHs for the late-forming and early-forming galaxies (as
labeled).  When comparing the two SFHs we can see that the SFH of the
$z_{50} > 2.9$ sub-sample has the majority of mass formed more rapidly
with a steeper decline, while the SFH of the $z_{50} < 2.9$ subsample
has a more gradual decline in SFR (with more star formation in the
recent past). \label{fig_sfh}}
\end{figure*} 
\subsection{Modeling the Stellar Populations and Star-Formation
  Histories}\label{section_modeling} 

To constrain the stellar population parameters of our galaxies we build on our forward modeling technique
described in \cite{estr19}, previously applied solely to WFC3/G102
grism data.
%
%
We use Flexible Stellar Population Synthesis 
(FSPS) models 
%
%
\citep[e.g.,][]{conr10}, using a combination of MILeS and
BaSeL libraries and assuming a Kroupa initial mass
function \citep{krou01}, to fit our SEDs.  

We have updated our methodology to use the dynamic nested sampling
algorithm engine from Dynesty \citep{spea19}. This allows us to model
additional parameters, and take advantage of improvements in
computational speed and parallelization.  Dynesty allows us to include
additional (nuisance) parameters to handle possible systematics which
arise when fitting the two spectroscopic data sets (deep G102 and
shallow G141 grism spectra) and broadband photometry simultaneously.
We include a parameter allowing for an additional linear slope applied
to the grism data (to account for corrections to the contamination
subtraction). We also introduce parameters to account for correlated
noise terms in the grism data described in \cite{carn19}. Our methods
will be described fully in a future paper (V.~Estrada-Carpenter et
al. in prep), where we will apply this method to the full CLEAR sample
to study the co-evolution of star-forming and quiescent galaxies.

Here we applied this method to all the quiescent galaxies in our parent sample (Section~2.1).  We use the WFC3 G102 + G141 data, and broad-band photometry (see Section 2).  

In this study we focus on the SFHs of quiescent galaxies at $0.7 <
z_{grism} < 2.5$. We adopt a ``non-parameteric" SFH parameterization
\citep{leja19}, which include parameters to describe the SFR in 10
time bins and allows for much greater flexibility in the SFHs.
\editone{The time bins are wider at larger look-back times (further in the past), except for the last (oldest) time bin, which is slightly smaller to allow for more dynamic range in the SFH. We allowed the time spanned by the full SFH to vary (however the fractional amount of time spanned in each time bin is fixed, see discussion in \citeauthor{leja19}).}

%
%
%
Our full stellar population models have 23 fitted
parameters: metallicity ($Z$), age, SFH (10 total
parameters), redshift, dust attenuation
(assuming a Milky Way model \citep{card89}), stellar mass
($\log(M_*/M_\odot)$), and 8 nuisance parameters (1 tilt parameter
and 3 correlated noise parameters for each of the two grism spectra). The
choice of prior on the SFH is important (as each prior has its own 
systematics), and should be motivated by properties of the sample. 
We use the continuity prior for our SFHs, as this has the effect of 
weighting towards SFHs that evolve more
smoothly \citep[see discussion in Section 2.2.3,][]{leja19}. 

\editone{We then applied this method to all the quiescent galaxies in our parent sample (defined in Section 2.1) using the WFC3 G102+G141 grism data and the broad-band photometry in these fields (see Section 2).  For each galaxy, we derive posteriors on each parameter in the model.  }
To generate constraints on our SFHs, we randomly draw from
the posteriors generating 5000 realizations of the SFH,
we then derive the median SFH and 68\%-tile range. \edittwo{
Figure~\ref{fig_data} shows examples of fits and constraints on the
SFHs for three galaxies in our sample.  }
For each galaxy, we show 1000 individual SFH draws, the median SFH,
and the 68\%-tiles.
Each case in this figure illustrates galaxies with qualitatively
different SFHs, including one galaxy with evidence of early formation
and rapid quenching (GSD 39170), one with evidence for early
formation with a slowly declining SFR (GND 21156),
one with evidence for a early, nearly constant SFR, followed by
slow quenching (GSD 40862), and \editone{one with what is possibly a burst of star-formation at 
a look-back time of $\sim$ 0.5 Gyr (GSD 24569)}.  These are characteristic of the galaxies
in our sample. \edittwo{ In addition, we provide with this
  \textit{Paper} an interactive appendix with the fits and constraints
  on all the galaxies in our sample, see the information and
  hyperlink in Appendix~\ref{sec_appendix_online}.}

We define the ``formation'' redshift, $z_{50}$,  of a galaxy by
integrating the SFH to the redshift where the galaxy had formed 50\%
of its stellar mass.
%
%
We define the 68\%-tile on $z_{50}$ from the SFH using the highest
density region \citep[the smallest region that contains 68\% of the
probability density,][]{bail18}.  The constraints on
\zfif\ are  illustrated for the three galaxies in
Figure~\ref{fig_data}.  
%

\begin{figure}[t]
\epsscale{1.2}
\plotone{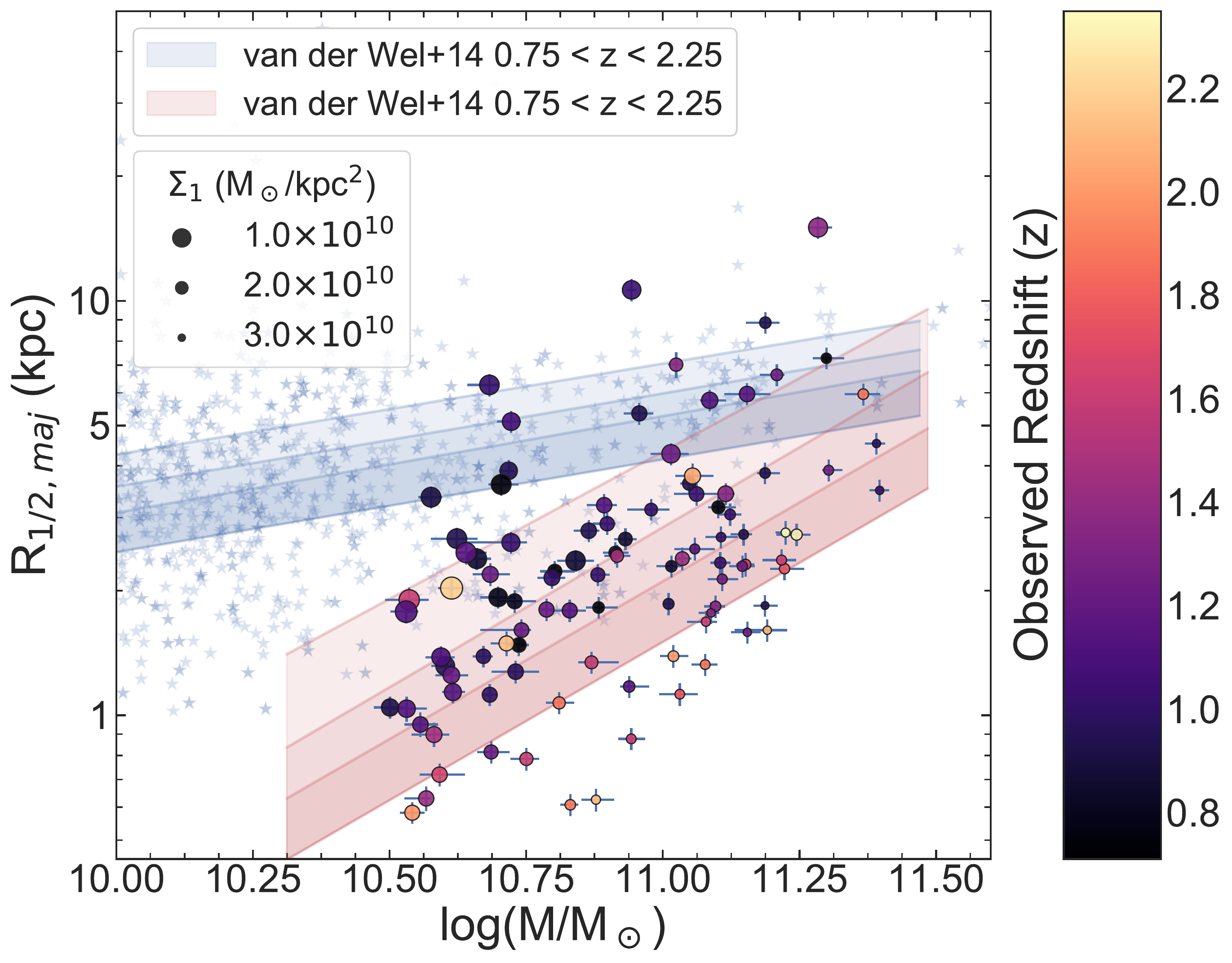}
\caption{\editone{Size mass relation for the $0.7 < z_\mathrm{grism} <
    2.5$ sample.}  The sizes of the points are 
scaled by their $\Sigma_1$ values, and their colors are scaled
by their redshift \edittwo{(star-forming galaxies in the CLEAR sample}
are shown 
as blue stars with no scaling). Size mass relations for star-forming (blue)
and quiescent (red) galaxies from \cite{vanw14} are shown. These span a range
from \editone{$0.75 < z < 2.25$ where the shading becomes darker with
increasing redshift. Following the results of the simulations of \cite{haus07}, 
we add a 6\%\ systematic error in quadrature to the $R_{1/2}$ values
to account for flux-dependent modeling uncertainties.
}
\label{fig_size_mass}}
\end{figure} 

To understand to what extent the galaxy photometric or spectroscopic
features are driving these differences in formation redshift, we 
inspected a subsample of galaxies at redshifts 
0.9 $<$ $z_{grism}$ $<$ 1.1.   We limit our sample to this
redshift interval so that our SFHs will have similar look-back times
and the data will have similar features present in the spectra.  We
then split this sample into ``early'' forming galaxies (\zfif\  $>$
2.9) and ``late'' forming galaxies (\zfif\ $<$ 2.9), normalize the
data at rest-frame $6000-6500$~\AA, and stack them (Figure~\ref{fig_sfh}). 
%
%

Figure~\ref{fig_sfh} shows that the ``late'' forming galaxies exhibit
a flux excess at $\lambda < 5500$~\AA, which increases into the
rest-UV. The gradient is largest around the 4000~\AA-break in the
ratio of the grism data (around the G+H$\gamma$ feature), implying
younger stellar populations exist in the ``late-forming'' $z_{50} <
2.9$ subsample. This is borne out in an inspection of other features
as well. For example, the ratio shows ``negative'' fluctuations at the
locations of all the Balmer lines (H$\alpha$, H$\beta$, H$\gamma$, and
(possibly) H$\delta$).  This is consistent with the differences in the 
subsamples being stellar populations with 
ages of $\lesssim 1$~Gyr, where we expect such
absorption to be strongest (i.e., dominated by A-type stars).
Furthermore, the stacked SFHs of the subsamples (inset panel in Figure
\ref{fig_sfh}) show that the ``early'' forming galaxies have high SFRs
at early times, peaking at $z \gg 4$, followed by a
relatively steep decline. In comparison the ``late'' forming galaxies
show more extended star formation that peaks at $z \sim 2.5-3$
followed by a gradual decline. We conclude the excess flux density
in the data at rest-frame UV/blue wavelengths drive the fits to
require more recent star formation in the ``late'' forming galaxies
compared to the ``early'' forming galaxies.  

\subsection{Measuring Compactness} 

We parameterize galaxy compactness using the stellar mass density
within 1~pkpc (proper kpc),  $\Sigma_1$
\citep[e.g.][]{fang13}. $\Sigma_1$ has advantages for quantifying
compactness as it uses information about the total
surface-brightness profile and is less sensitive to uncertainties and
correlations in quantities such as Sersic index ($n_s$) and effective
radius, $R_{1/2}$, \citep{lee18}.  Furthermore, using $\Sigma_1$ is
less susceptible to color gradients that can impact quantities such as
the half-light radius \citep[e.g.,][]{szom13,sues19}.  

We define $\Sigma_1$ using
the measured (total) stellar mass and the measured surface-brightness profile,
\begin{equation}
\Sigma_1 = \frac{\int _0 ^{1~\mathrm{kpc}}\ I_X(r)\ 2 \pi r\ dr}{\int _0 ^{\infty}\ I_X(r)\ 2 \pi r\ dr} 
\frac{L_\mathrm{GALFIT}}{L_\mathrm{phot}}\ \frac{M_{*}}{\pi(1\ \mathrm{pkpc})^2}
\end{equation}
where $I_X(r)$ is the S\'ersic profile measured in bandpass $X$ from
\citet[]{vanw14}. The ratio of the integrals
measures  the fraction of light within 1 pkpc compared to the total
light.  The ratio  $L_\mathrm{GALFIT}/L_\mathrm{phot}$ corrects for
differences in the total magnitude from the GALFIT fits and the
measured total photometry.   $M_{*}$ is the total stellar mass from
our fits.    To account for changes in rest-frame wavelength, we use
the surface-brightness profile measured in the WFC3/F125 ($J_{125}$)
bandpass for galaxies at $z_{grism} < 1.5$ and those measured in the
WFC3/F160W ($H_{160}$) bandpass for galaxies at $z_{grism} > 1.5$
\citep[see][]{vanw14}. 

\begin{figure*}[t]
\epsscale{1.1}
\plotone{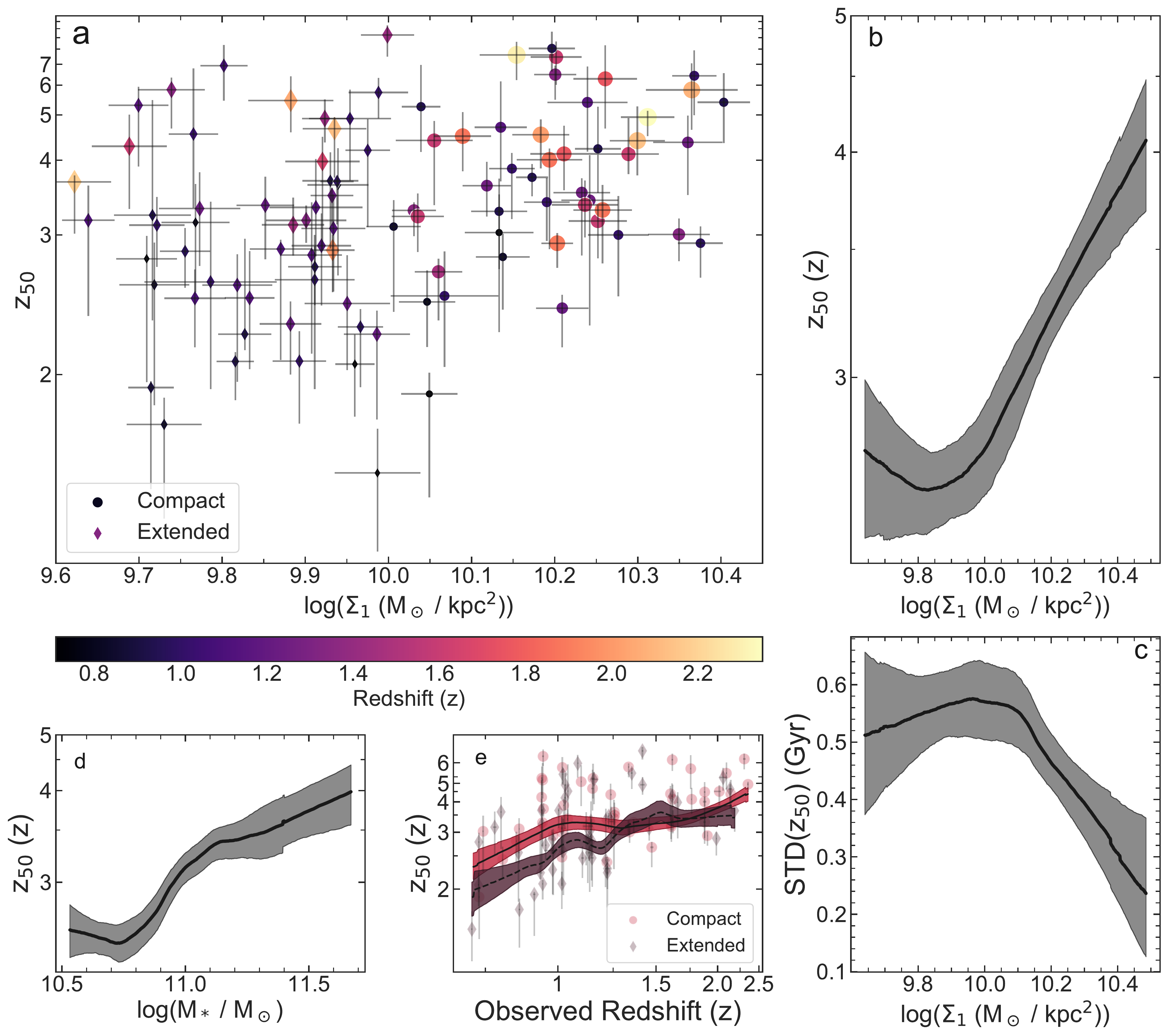}
\caption{Relationship between formation \editone{redshift \zfif\ (the
redshift by when 50\% of the stellar mass had formed), the observed
redshift $z_{grism}$, and $\Sigma_1$ (the stellar mass surface density
within 1~(proper) kpc).} (a) shows \zfif\ as a function of
$\log(\Sigma_1)$ for the quiescent galaxies in our sample.  Galaxies
with $\log \Sigma_1 / (M_\odot\ \mathrm{kpc^{-2}})) > 10$ ($<10$)  are
shown as circles (diamonds).  \editone{The  color and size of the all
points scales with increasing $z_{grism}$}. Galaxies with $\log
\Sigma_1 / (M_\odot\ \mathrm{kpc^{-2}}) < 10$ span a larger  range of
$\zfif$. Galaxies with $\log \Sigma_1 / (M_\odot\ \mathrm{kpc^{-2}}) >
10$ favor higher formation redshifts of $\zfif > 3$. \editone{(b)
shows the change in \zfif\ as a function of \Lsig\ using a LOWESS
algorithm \editone{with bootstrapping to estimate the 68\% confidence
region}. (c) shows the scatter  in \zfif\ as a function of \Lsig\
(using LOWESS).   Galaxies with higher $\Sigma_1$ tend towards higher
\zfif\ with lower scatter.}
%
%
 (d) shows the change in \zfif\ as function of
 $\log(M_*/M_\odot)_{grism}$ using LOWESS.
Higher mass galaxies tend
towards higher \zfif, \editone{though this relation is less steep while there is a continued rise between 
\zfif\ versus the stellar-mass surface density, $\Sigma_1$.}  (e) shows the
formation redshift, \zfif, against the observed redshift.  Galaxies
with \Lsig $/ (M_\odot \mathrm{kpc}^{-2}) > 10$ ($<10$) are indicated
by red (purple) points, using a LOWESS algorithm to show the
trend. \editone{We see here that more compact  galaxies (i.e., with
higher $\Sigma_1$) tend to have higher \zfif, particularly
  for $z \lesssim 1.25$.}
\label{fig_sigma_1}}
\end{figure*} 

Figure \ref{fig_size_mass} shows the relation between the effective
radii (major axis) and stellar masses for the galaxies in
our samples (i.e., the size-mass relation). 
The size (hue) of the data points are scaled by the
$\Sigma_1$ ($z_\mathrm{grism}$) values.
Galaxies with the largest $\Sigma_1$ (highest
compactness) tend to sit at the high-mass/low-size end of the
distribution. This is to be expected as
$\Sigma_1$ is derived based on both the stellar mass and the
surface-brightness profiles (which depends on $R_{1/2}$).
Furthermore, we see no significant correlation between
$\Sigma_1$ and $z_\mathrm{grism}$ (the redshift measured from our
WFC3/G102 + G141 grism data):  galaxies with the highest (and lowest)
$\Sigma_1$ among our sample span a range of observed redshift.

\section{Results \label{sec_res}} 

\subsection{Compact Galaxy Formation}

Figure~\ref{fig_sigma_1}a shows the main result of our study: galaxies
with compact stellar mass surface brightnesses \Lsig$/(M_\odot\
\mathrm{kpc}^{-2})$ $>$ 10.25 favor almost exclusively earlier \zfif\
values ($\zfif > 3$).    Among the subsample of objects that fall in
this ``ultra-compact'' region (defined by \Lsig$/(M_\odot\
\mathrm{kpc}^{-2})$ $>$ 10.25) there are no examples of galaxies with
lower formation redshifts. Recall that all the galaxies in
Figure~\ref{fig_sigma_1} are classified as ``quiescent'' using the
same ($UVJ$) selection criteria, and have no explicit selection by
galaxy morphology.  Therefore, it is striking that the SFHs of the
most compact galaxies, as defined by $\Sigma_1$, disfavor low
formation redshifts, $z_{50}$. \editone{We find the same conclusion if we
define ``quiescent'' using a selection of sSFR $<10^{-11}$ yr$^{-1}$.}

Figure~\ref{fig_sigma_1}b reinforces the observation that the stellar
mass surface density, $\Sigma_1$,  is related to the formation epoch
\zfif.  Here we smooth \zfif\ as a function of $\Sigma_1$ using 
locally weighted scatterplot smoothing (LOWESS) and see that 
the relationship monotonically  rises as a function
of compactness. Figure~\ref{fig_sigma_1}c shows that the
standard deviation in \zfif\ of the sample changes as a function of
compactness (using LOWESS as well), reinforcing that the
dynamic range of \zfif\  is dependent on $\Sigma_1$. 

Figure~\ref{fig_sigma_1}d shows \zfif\ as a function of
$\log{(M_*/M_\odot)_{grism}}$ using LOWESS. Galaxies with higher
stellar masses do tend to have earlier \zfif\ than lower mass
galaxies, though this relationship seems to plateau for $\log
(M_\ast/M_\odot)_{grism} \gsim 11$, \editone{where $\zfif$ increases more slowly for increasing stellar mass ($dz/d\log(M) \simeq 1.2$).  In contrast, there is a steeper relation between $\zfif$ and $\Sigma_1$:  $dz/d\log (\Sigma_1) \simeq 2.7$ for $\log (\Sigma_1) /(M_\odot\ \mathrm{kpc}^{-2}) > 10$.  Therefore, while $z_{50}$ is correlated with both stellar mass and stellar-mass surface density, the trend is stronger with the latter. }

The preference for early formation of the most compact galaxies does
not appear to be due to redshift selection effects.
The galaxies in our sample do
span a range of \textit{observed} redshift, and if there is a correlation
between observed redshift and formation redshift, then this could
account for our findings.  Figure~\ref{fig_sigma_1}e shows
this is not the case. The distribution
of $z_{50}$ for quiescent galaxies shows 
that the more compact quiescent
galaxies tend towards higher formation redshifts, $z_{50}$. 
\editone{This separation is most pronounced for redshifts $z\lsim
  1.25$ (Figure~\ref{fig_sigma_1}e).  At higher redshifts, $z\gsim
  1.25$,  there
  is no difference in the distribution of $z_{50}$ and observed
  redshift.   A larger sample of high redshift galaxies would be
  necessary to see if the separation observed at $z \lsim 1.25$
  extends to higher redshift.  } 

We also considered (and rejected) the possibility that our $\Sigma_1$
values are dependent on color gradients.  For the sample with $z_{grism} <
1.5$ we recalculated $\Sigma_1$ using the $H$--band surface-brightess
profile fits \citep[from][]{vand14}. The $\Sigma_1$ values change by
$<$5\% implying the stellar surface densities for $r < 1$~pkpc are
robust to color gradients observed to affect the effective radii of
galaxies \citep[e.g.,][]{szom13,sues19}. 

There is also no apparent bias between $\Sigma_1$ and SFH.  The
derivation of SFH constraints and the measurement of $\Sigma_1$ are
almost entirely independent. The stellar-mass surface density stems
from the morphological surface brightness profile. While the
morphological profile can affect the spectroscopic resolution of the
\hst/WFC3 grisms \citep[galaxies with more compact morphologies have
higher resolution, see][]{vand11, estr19}, this is mild for the
galaxies in our sample (the spectroscopic resolution changes by a
factor of $\lesssim 2$).  Moreover, as shown in Figure~\ref{fig_sfh}
the differences in the spectral energy distributions of ``early'' and
``late'' forming galaxies extends through the full broadband
photometry.  Therefore, our results show that ultra-compact massive
quiescent  galaxies had at least 50\% of their stellar-mass in-place
at  $z \gsim 3$.
%


\begin{figure*}[t]
\epsscale{1.1}
\plotone{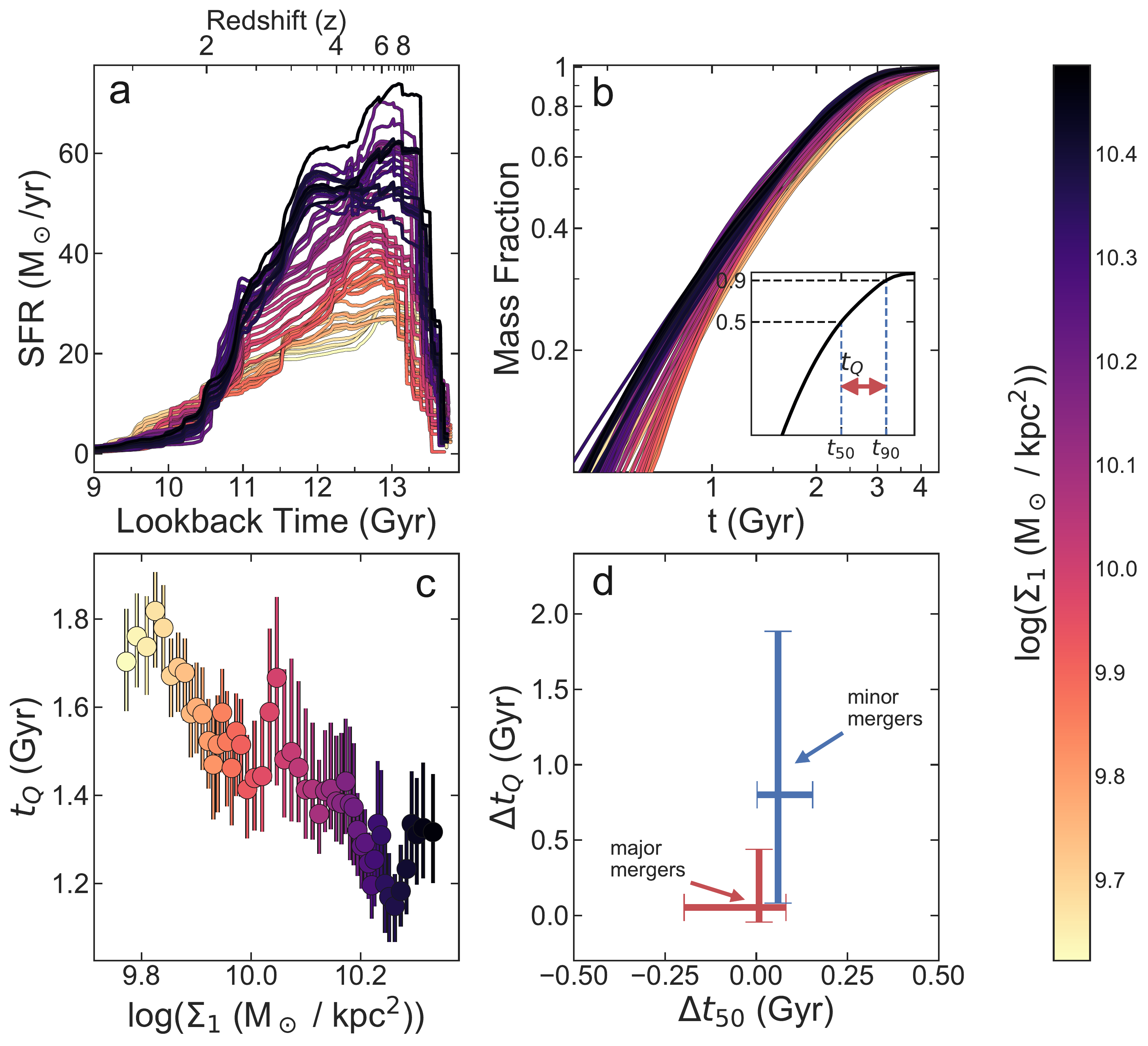}
\caption{The relation between SFHs, quenching times, and stellar mass
surface density ($\Sigma_1$) for  ``early-forming'' galaxies ($z_{50}
> 2.9$). (a) shows a the mean SFH for galaxies stacked as a
function of \Lsig in bins of 0.2 dex. (b) shows the cumulative fraction of stellar mass
formed. Both (a) and (b) show that galaxies with higher \Lsig\ form
more stellar mass earlier with higher peak SFRs, and
experience a more rapid decline in their SFR compared to galaxies with
lower $\Sigma_1$. (c) shows the quenching timescale ($t_Q$) defined as the
time between when the galaxy had formed 50\% and 90\% of its stellar
mass, as a function of $\Sigma_1$, with error bars derived from bootstrapping.
Galaxies with higher $\Sigma_1$ have
shorter quenching times. (d) shows the effects of mergers on the SFH
timescales.  We randomly merged simulated galaxies and measured the
change in $t_Q$ and $t_{50}$ from major-mergers (mass ratios $>$1:4;
red) and minor-mergers (mass ratios $<$1:10; blue).  The error bars
show the inter-68\%-tile scatter (68\% of the simulations fall in this
range). 
\label{fig_dt}}
\end{figure*} 

\subsection{Quenching Timescales}

The main question that arises from our results is what specific
properties of galaxies drive the lack of ultracompact quiescent
galaxies with \zfif\ $<$ 3?   There are measurable
differences in the SFHs of galaxies as a function of \Lsig.
Figure~\ref{fig_dt}a shows the mean SFHs for all galaxies with $z_{50}
> 2.9$ as a function of \Lsig. In this figure, each curve  corresponds
to the mean SFH within a 0.2 dex bin of \Lsig centered on the value
illustrated by the color bar.  The peak SFR increases with \Lsig, and
the shape of the SFH varies with $\Sigma_1$. Galaxies with lower $\Sigma_1$
have a flatter overall shape to their SFHs with a more gradual decline
in SFR.

These differences in the SFH with $\Sigma_1$ for the $z_{50} > 2.9$
galaxies are evident in the time evolution of the cumulative fraction
of stellar mass, illustrated in Figure \ref{fig_dt}b.  Qualitatively,
both Figures~\ref{fig_dt}a and B  show that galaxies with the largest
stellar-mass surface densities ($\Sigma_1$) formed their stellar mass
more rapidly and at earlier times compared to galaxies with lower
$\Sigma_1$.

We can quantify this point by defining a ``quenching timescale'',
$t_Q$,  as the time (in Gyr) needed for the SFH to form 50\% of the mass to
90\% of the mass ($t_Q \equiv t_{50} - t_{90}$, illustrated in
Figure~\ref{fig_dt}b). Figure~\ref{fig_dt}c shows the $t_Q$ values as
a function of $\Sigma_1$ for the galaxies with $z_{50} > 2.9$,
\editone{with errors derived from bootstrapping}. There
is an apparent (anti-)correlation between stellar-mass surface
density, $\Sigma_1$ and quenching timescale, $t_Q$. Galaxies at lower
$\Sigma_1$ ($\log \Sigma_1/(M_\odot\ \mathrm{kpc^{-2}}) < 10.0$) have $t_Q
\gtrsim 1.4$ Gyr.  In contrast, galaxies with the highest stellar mass
surface density ($\log \Sigma_1/(M_\odot\ \mathrm{kpc^{-2}}) > 10.2)$ have
shorter quenching times, with $t_Q \simeq 1.2 - 1.4$ Gyr. 
The faster quenching timescales of the
ultra-compact ($\log \Sigma_1/(M_\odot\ \mathrm{kpc^{-2}}) > 10.25$)
sub-sample indicates that these
galaxies have an overall more rapid SFH with faster quenching (shorter
$t_Q$). 
%
%

\section{Discussion}

The main finding of our study of the broad-band photometry and
\hst/WFC3 grism spectroscopy of quiescent galaxies at $0.7 < z_{grism} < 2.5$
is that they show evidence for a relation between their SFHs (e.g.,
formation redshifts, $z_{50}$), and their morphologies parameterized
by their stellar-mass surface density within 1 (proper) kpc,
$\Sigma_1$ (Figure~\ref{fig_sigma_1}).  Galaxies with high $\Sigma_1$,
($\log \Sigma_1 / (M_\odot\ \mathrm{kpc^{-2}})$ $>$ 10),  typically
have higher \zfif\ values, where ultra-compact galaxies with $\log
\Sigma_1/ (M_\odot~\mathrm{kpc^{-2}}) > 10.25$ all have $z_{50}>2.9$.
They are "early forming".   Less compact galaxies ($\log \Sigma_1 /
(M_\odot\ \mathrm{kpc^{-2}}) < 10$) on the whole have lower average
formation redshifts, but they span a wide range, $z_{50} \sim 1-8$.  
Galaxies with higher $\Sigma_1$ show SFHs that have higher peak SFRs at earlier
times, with more rapid quenching times.
Both the shorter quenching times and earlier
$z_{50}$ values for ultracompact galaxies 
suggests that these properties are a symptom of
the physics related to galaxy quenching.  
%

\subsection{Our Results in Context}\label{section_context}

Our findings reinforce some earlier studies
\citep[e.g.,][]{tacc17,will17,lee18,wu18}, which found evidence of older ages
in compact galaxies when compared to extended galaxies.  Likewise,
some studies found that compact galaxies also show evidence of
quenching more rapidly \citep{barr13,barr17,nogu19}.   In addition,
many of our galaxies have relatively high formation redshifts ($z_{50}
\gtrsim 5$), suggesting they may be the descendants of quenched
galaxies recently identified at high redshift ($z \gtrsim 3$)
\citep[e.g.,][]{spli14,stra14,mars15,glaz17,schr18b,tana19,forr20,vale20}.  Indeed, quiescent
galaxy candidates at $3 < z < 4$ have very compact sizes
\citep{stra15}, consistent with idea that these galaxies have high
$\Sigma_1$ and could be among the  progenitors of the early-forming
galaxies in our sample. 

Our conclusions depend on the reliability of the SFH
constraints.  To gauge this, we compared our results to other studies
of massive galaxies at similar redshifts. These broadly show a
correlation between stellar mass, and shorter, more intense formation
periods at higher redshift \citep{paci16,schr18b,carn19,mori19}.
%
%
%
Our results are in line with these studies, where we do see a trend
between $z_{50}$ and stellar mass (Figure~\ref{fig_sigma_1}d).

\subsection{Implications for the Evolutionary Paths of Quiescent Galaxies}

A key new result is evidence for a trend
between (increasing) stellar-mass surface density, $\Sigma_1$, and
(higher) formation redshift, $z_{50}$, for galaxies in our sample.  The
``early-forming'' galaxies ($z_{50}{>}2.9$) have quenching timescales
($t_Q$) that decrease with increasing $\Sigma_1$
(Figure~\ref{fig_dt}).    
%

\subsubsection{On the Origin of Early-Forming Galaxies with High $\Sigma_1$}

The origin of galaxies with high stellar mass surface
density  ($\log \Sigma_1 / (M_\odot\ \mathrm{kpc^{-2}}) > 10.25$)  at
higher $z_{50}$ is expected as a consequence of the gravitational
collapse of galaxies at high redshift (to overcome the cosmic background density,
e.g., \citealt{well15,lilly16}).  
%
%
\cite{well15} show simulations where the earliest forming
quiescent galaxies achieve central stellar densities of $\log
\rho(<1~\mathrm{kpc}) / (M_\odot~\mathrm{kpc^{-3}}) > 10$ by $z > 5$.

Other explanations for high $\Sigma_1$ seem less likely. Mergers
seem insufficient as major mergers are expected to leave $\Sigma_1$
roughly unchanged, while minor mergers can \textit{decrease
$\Sigma_1$} (see \citealt{beza09} and below).  These 
galaxies are also unlikely to be the product of the
``compaction'' \citep[e.g.,][]{deke09,barr13} or from
gas-rich mergers \citep[e.g.,][]{well15}. These processes should be more
frequent at later times, where we do not
observe any galaxy with $\log \Sigma_1 / (M_\odot\ \mathrm{kpc^2}) >
10.25$ and $z_{50} < 2.9$. Compaction events or major gas-rich mergers
for these galaxies are either rare or are unable increase the mass 
surface density to $\log  \Sigma_1/
(M_\odot\ \mathrm{kpc^2}) > 10.25$.
%

\subsubsection{On the Origin of Early-Forming Galaxies with Low $\Sigma_1$}\label{section_earlylow}

There are two possibilities to explain the existence of galaxies with both high formation redshift ($z_{50} >
2.9$) and lower $\Sigma_1$ ($\log \Sigma_1 / (M_\odot\
\mathrm{kpc^{-2}}) < 10.25$).  These galaxies could form with intrinsically lower
$\Sigma_1$, but this is unexpected given the arguments above.  Alternatively,
these galaxies may form the bulk ($\gtrsim 50\%$) of their stellar
populations at $z > 2.9$ with high $\Sigma_1$, but then experience
evolution that reduces $\Sigma_1$.   This could
come from the adiabatic expansion through mass losses from late stages
of stellar evolution \citep[e.g.][]{vand14,barr17}.  However, this
becomes more efficient at later times, and there is only $\sim$4 Gyr
between $z\sim 2.9$ and $z\sim 1$ for this to
manifest.  \citet{well15} show the central density within 1~kpc of
an early-forming compact quiescent galaxy at $z\sim 5$ declines by
$\lesssim 0.1$~dex by $z\sim 2$.  Furthermore, it is unclear why this affects only some of the
ultracompact galaxies when stellar evolution should impact all.  

One important clue comes from the correlation between (longer) quenching times, $t_Q$, and (decreasing) $\Sigma_1$.  The early-forming galaxies (with
$z_{50} > 2.9$)  and lower $\Sigma_1$
($\log \Sigma_1 / (M_\odot\ \mathrm{kpc^2}) < 10.25 $) have longer
quenching times, compared to galaxies with  $\log \Sigma_1 / (M_\odot\
\mathrm{kpc^2}) > 10.25$ (see Figure~\ref{fig_dt}c). 
%
%
%
%
%
An explanation for this correlation is that all early-forming massive
galaxies begin with high $\Sigma_1$. Galaxies then experience a unique
assembly history, where the frequency, orbital configuration, and
distribution of mass-ratios of mergers and accretion events dictates
the change in $\Sigma_1$.  \citet{beza09} show minor mergers
(mass ratios $\lesssim$1:10) can decrease $\Sigma_1$, while
major mergers (mass ratios greater than $\gtrsim$1:4) leave $\Sigma_1$
mostly unchanged. 

Minor mergers involve high-mass and low-mass systems.  The latter have more prolonged SFHs (see
Section~\ref{section_context}).   We tested how this would impact the
formation times $t_{50}$ (corresponding to $z_{50}$) and the quenching
time $t_Q$ using a series of simulations.  We simulated galaxy SFHs as ``delayed-$\tau$'' models \citep[e.g.,][]{estr19} using the correlations between SFH and stellar mass (see above).  We then randomly ``merged'' galaxies of different mass ratios, summing their SFHs to simulate the effects of mergers on the integrated SFH.  Figure~\ref{fig_dt}d shows the results.  Major mergers have little effect on neither $t_{50}$ nor $t_Q$, which
change by $<$0.25~Gyr (recall that $t_Q \equiv t_{90} - t_{50}$).  Minor mergers, on the other hand, have little
effect on $t_{50}$ (change by $\lesssim$0.2~Gyr) but can extend the 
SFHs with an increase in $t_{90}$, making $t_Q$ longer with a
scatter of up to $\sim$ 2~Gyr.  Therefore, minor mergers provide a mechanism to increase the scatter in $t_Q$ with 
only a small change in $z_{50}$ (the redshift corresponding to
$t_{50}$), \textit{and} decrease $\Sigma_1$ \citep{beza09}, which is consistent with the observations. 

\subsection{\editone{On the lack of ``Early-Forming'' Galaxies at low-redshift}}

\begin{figure}[t]
\epsscale{1.2}
\plotone{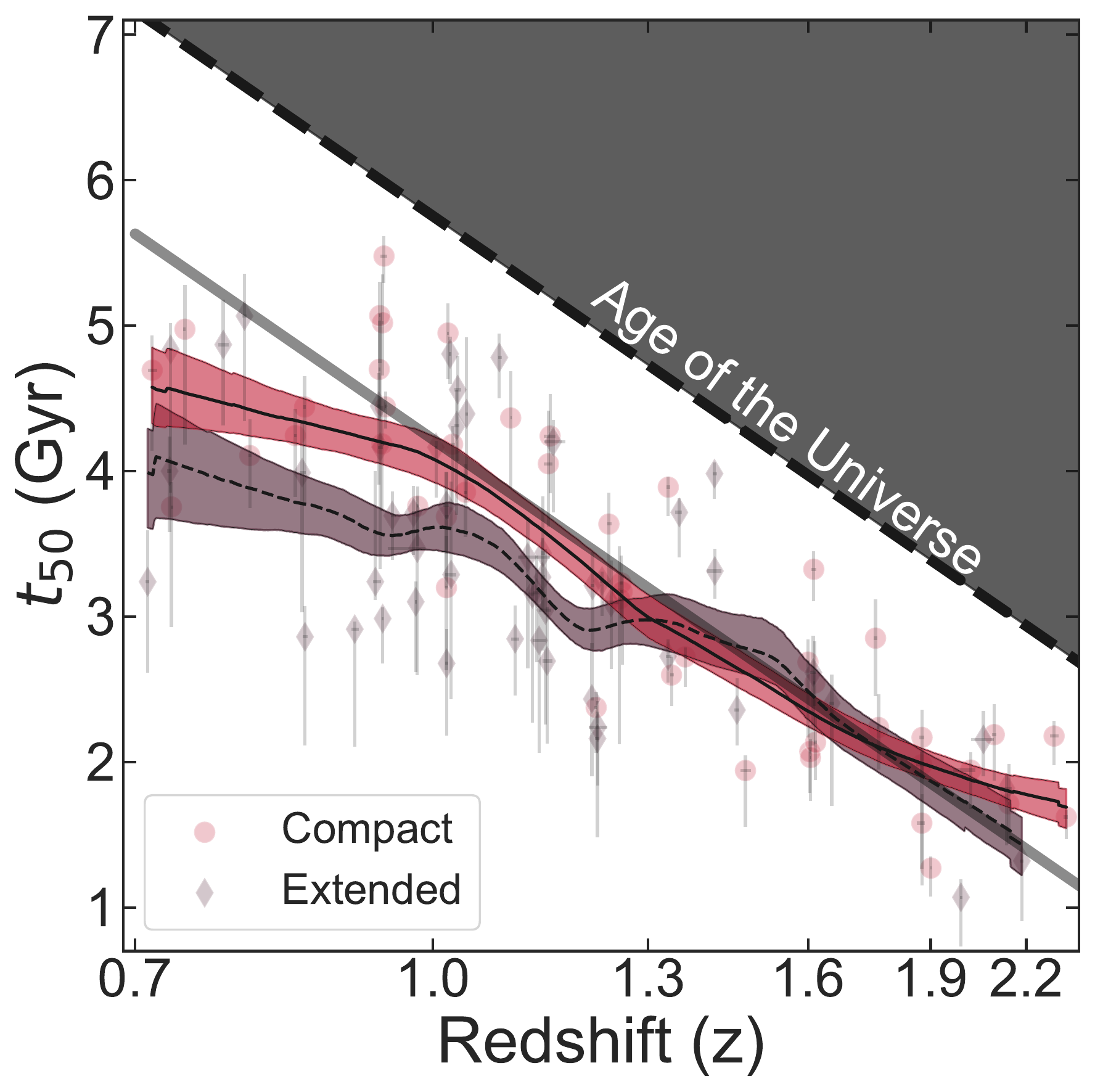}
\caption{\editone{The formation age ($t_{50}$) as a function of
    observed redshift, $z_\mathrm{grism}$ for the quiescent galaxy sample.
    The formation age is the lookback time from the observed redshift
    for a galaxy to its formation redshift, $z_{50}$, when it had
    formed 50\% of its stellar mass.  The symbols divide the sample
    into subsamples of compact (red circles,  $\log \Sigma_1 / (M_\odot\
    \mathrm{kpc}^{-2}) > 10$) and 
extended sources (purple diamonds,  $\log \Sigma_1 / (M_\odot\
    \mathrm{kpc}^{-2}) < 10$).    The solid swath tracks the trend for
    each subsample using a LOWESS algorithm
with bootstrapping.  The dashed diagonal line demarcates the  age of
the Universe at the observed redshfit, and the solid grey line shows
the age of the Universe minus 1.5 Gyr.  At high redshift, $z \gtrsim
1.25$ the galaxies' formation ages mostly track the age of the Universe
offset by $\sim$1.5 Gyr.  At lower redshifts the populations skew
toward more recent formation, but at different redshifts.  The
extended sample skews toward lower $t_{50}$ at earlier times ($z
\lesssim 1.25$) while the compact galaxies skew toward lower $t_{50}$
at later times ($z_{50} \lesssim 0.9$).}
\label{fig_t50}}
\end{figure} 

\editone{Figure~\ref{fig_sigma_1}e shows an absence of quiescent
  galaxies at lower observed redshifts ($z_{grism} < 0.9$)
and early formation times, $z_{50} > 4$.  We considered several reasons
that could explain this absence, some systematic to the data/analysis
and  others physical.  }

\editone{One potential systematic reason (which we ultimately reject)
  could be that galaxies at lower redshifts  lack (grism)
spectroscopic coverage in the rest-frame UV, and this could limit our
ability to constrain the current SFRs in those cases.   The WFC3 G102 grism
covers $>$0.8~$\mu$m, corresponding to $\gsim$4000~\AA\ in the
rest-frame for $z\sim 1$ galaxies.  We therefore tested if this could
limit our ability to identify objects with early star-formation at
these observed redshifts.  We simulated the spectral energy
distribution of a quiescent galaxy at $z =
0.8$ with early quenching, with $z_{50} = 8$.  We then perturbed the
photometry and grism data for this object by the measured
uncertainties, and repeated the model
fitting using our procedures applied to the real CLEAR galaxies.   In
this case we reliably recover this \zfif\ value, within a 68$\%$
confidence interval of $\pm$ 0.15 Gyr.
Therefore it appears that if galaxies at $z=0.8$ with $z_{50} \gg 4$
existed in our dataset we would identify them as such. }

\editone{One other systematic reason for the lack of objects with
$z_\mathrm{grism} < 0.9$ and $z_{50} \gg 4$ could be that the CLEAR
data sample a relatively small volume.   For example, the comoving
volume probed by our study is $\approx$8 times larger for $1 < z <
2.5$ than $0.7 < z < 1.0$, and these objects with early formation and
lower observed redshift \edittwo{may simply be rarer at these
redshifts.} Future studies using larger datasets should be able to test
this systematic more fully.}

    \editone{ Alternatively, the rarity of early-forming quiescent
galaxies ($z_{50} \gg 4$) at $z < 0.9$ could be indicative of how
these galaxies evolve.   Building off the discussion above
(Section~\ref{section_earlylow}) we expect that quiescent galaxies
grow in size through mergers, and this evolution depends on the
galaxies' individual assembly histories.  Our toy model argues these
mergers both  lower $\Sigma_1$ and \textit{decrease}  $z_{50}$ and
that the magnitude of both affects should grow with time.  We
therefore can predict that galaxies with early quenching observed at
lower redshift would have lower  (measured) $z_{50}$
and lower $\Sigma_1$ and this effect should become more pronounced with
decreasing redshift.} 

\editone{ Interestingly, our results support this
  interpretation. Figure~\ref{fig_t50} shows the relationship
  between formation \textit{age}, $t_{50}$, and observed redshift
  $z$.  Here, $t_{50}$ is the lookback time between the observed
  redshift and the formation redshift $z_{50}$ for each galaxy.
  In the figure we split our quiescent galaxies into samples of
  compact and extended based on $\log \Sigma_1 / (M_\odot\
  \mathrm{kpc}^{-2}) > 10$ or $< 10$, respectively (see Figure
  \ref{fig_sigma_1}). Both the compact and extended galaxies have
  similar evolution at $z \gtrsim 1.25$:  their quenching time, $t_{50}$
  is (on average) roughly 1.5~Gyr delayed from the Big Bang
  \citep[and this is consistent with the currently earliest known
galaxies with older stellar
populations,][]{glaz17,schr18,forr20,vale20}.}

\editone{
However, the trend between observed redshift and $t_{50}$ for the
extended and compact quiescent galaxies diverges at observed redshifts
of $z \lesssim 1.25$.    Here the extended galaxies show lower
$t_{50}$ (at fixed observed redshift) compared to the compact
galaxies.  This could be a result of the hypothesis that the extended
galaxies have experienced more frequent growth due to minor mergers,
causing a faster decrease in $\Sigma_1$ (making them ``extended'' as 
described in Section \ref{section_earlylow}) and
in $t_{50}$.  However, the subsample of quiescent galaxies at $ z<
0.9$ in our sample remains small, and larger samples will be needed to
confirm these trends.}

\section{\editone{
 Conclusions}}

\editone{
In this paper we constrain the star-formation histories
of quiescent galaxies at $0.7 < z < 2.5$ and correlate these with
galaxy masses and morphologies, using ``non-parametric'' star-formation
histories and a nested sampling algorithm.  We derived constraints for the
formation and quenching timescales for a sample of nearly 100 quiescent
galaxies with deep \hst\ grism spectroscopy and photometry
from the CLEAR (CANDELS Lyman$-\alpha$ Emission at Reionization)
survey.   } \edittwo{In addition to the results presented here, we
  provide in   Appendix~\ref{sec_appendix_online} a
  hyperlink to, and a description of, an online appendix that contains similar fits and
  information for all the galaxies in our sample.  Our conclusions
  from this study are as follows.}

\begin{enumerate}

  \item  The galaxy formation redshifts, $z_{50}$ (defined as the point
where they had formed 50\% of their stellar mass) range from
$z_{50}\sim 2$ (shortly prior to the observed epoch) up to $z_{50}
\simeq 5-8$. We correlate the formation redshifts with the
stellar-mass surface densities, $\log \Sigma_1 / (\Mkpcsq)$,
where  $\Sigma_1$ is the stellar mass within a 1~pkpc (proper kpc).  
\item Quiescent galaxies with the highest stellar-mass surface density,
$\Sigma_1 > 10.25$,  show a \textit{minimum} formation redshift: all such objects in our
sample have $z_{50} > 2.9$.
\item Quiescent galaxies with lower surface
density, $\log \Sigma_1 / (M_\odot\ \mathrm{kpc}^{-2}) = 9.6 - 10.25$,
show a range of formation epochs ($z_{50} \simeq 1.5 - 8$), implying
these  galaxies experienced a range of formation and assembly
histories. 
\item We argue that the surface density threshold
$\log\Sigma_1/(M_\odot\ \mathrm{kpc}^{-2})>10.25$ uniquely identifies
galaxies that formed in the first few Gyr after the Big Bang

\end{enumerate}

\editone{
}    


\editone{We then discuss the implications this has for galaxy formation and
quenching. Based on our data, the ultracompact quiescent galaxies ($\log \Sigma_1 /
(M_\odot\ \mathrm{kpc^{-2}}) > 10.25$) appear to identify galaxies with
early formation ($z_{50} > 2.9$) and a lower fraction of mergers (at
the time they are observed, see Section~\ref{section_earlylow}).  
If these exist in the present Universe, they could be compact cores
of galaxies. It could be instructive to
identify objects with high density cores, and study their ages,
abundances, and gradients.   Additional simulations would be useful both to understand the
formation and the evolution of these galaxies, and if later time 
processes (such as adiabatic expansion or mergers) destroy them. 
  Alternatively, it may be that examples of these objects still exist in the present-day
Universe.  If so, the most compact passive galaxies today may host the
oldest stellar populations and be the remnant of these bygone eras.}

\editone{We favor the conclusion that stochasticity in the 
mergers/accretion history of lower-mass early-forming galaxies
($z_{50} > 2.9$) explains the relation between the quenching timescale 
and stellar mass surface density:  the lower  $\Sigma_1$  ($\log
\Sigma_1 / (M_\odot\ \mathrm{kpc^{-2}}) \lesssim 10.1$) and longer
quenching times ($t_Q > 1.4$~Gyr) of these galaxies is a result of
their history of (minor) mergers.}

\editone{The formation redshift, $z_{50}$  (or age, $t_{50}$) can be
reduced  through subsequent evolution  through minor mergers and this
can lead to both galaxies with high $z_{50}$ and lower stellar-mass
surface densities as well as account for the lack of observed galaxies
at $z \lesssim 0.9$ with early formation times (high $t_{50}$).  The
obvious caveat to this interpretation is that we have neglected the
contribution of ``progenitor bias'' \citep[see, e.g.,][]{vand99}
whereby newly quenched galaxies are continuously becoming
``quiescent'' at later times.  As the more recently-quiescent galaxies
will (by definition) have lower $t_{50}$ and likely have lower
$\Sigma_1$,  they can also contribute to the trend seen between
observed redshift and quenching time ($t_{50}$) in
Figure~\ref{fig_t50} \citep[though see,
e.g.,][]{deke09,barr13,well15}.   Ultimately, it is likely that both
the effects of early formation plus minor mergers and progenitor bias
are at work. This makes an interesting  prediction that spatially
resolved studies should see variations in the SFH (or possibly
abundance histories) as a function of galactic radius in these
galaxies. This may be testable with data from either the \textit{James
Webb Space Telescope} (\jwst) or 25--30~m-class telescopes.}

\acknowledgments We thank our colleagues on the CLEAR team for their
valuable conversations and contributions.  \editone{We also thank
Kartheik Iyer, Rob Kennicutt, Arjen van der Wel, Sandro Tacchella, and
Christina Williams for productive comments, feedback, suggestions, and
information. We also thank Mark Dickinson and Hanae Inami for
providing and assisting with the 24~$\mu$m catalog.  We are also
grateful to the anonymous referee whose comments and queries improved
the quality and clarity of this paper.}     VEC acknowledges support
from the NASA Headquarters under the Future Investigators in NASA
Earth and Space Science and Technology (FINESST) award
19-ASTRO19-0122, as well as support from the Hagler Institute for
Advanced Study at Texas A\&M University.  This work is based on data
obtained from the Hubble Space Telescope through program number
GO-14227.  Support for Program number GO-14227 was provided by NASA
through a grant from the Space Telescope Science Institute, which is
operated by the Association of Universities for Research in Astronomy,
Incorporated, under NASA contract NAS5-26555.  This work is supported
in part by the National Science Foundation through grants AST
1614668. The authors acknowledge the Texas A\&M University Brazos HPC
cluster and Texas A\&M High Performance Research Computing Resources
(HPRC, \url{http://hprc.tamu.edu}) that contributed to the research
reported here. 

\bibliography{Vince_compact_galaxies}{}

\begin{figure}
\epsscale{1}
\plotone{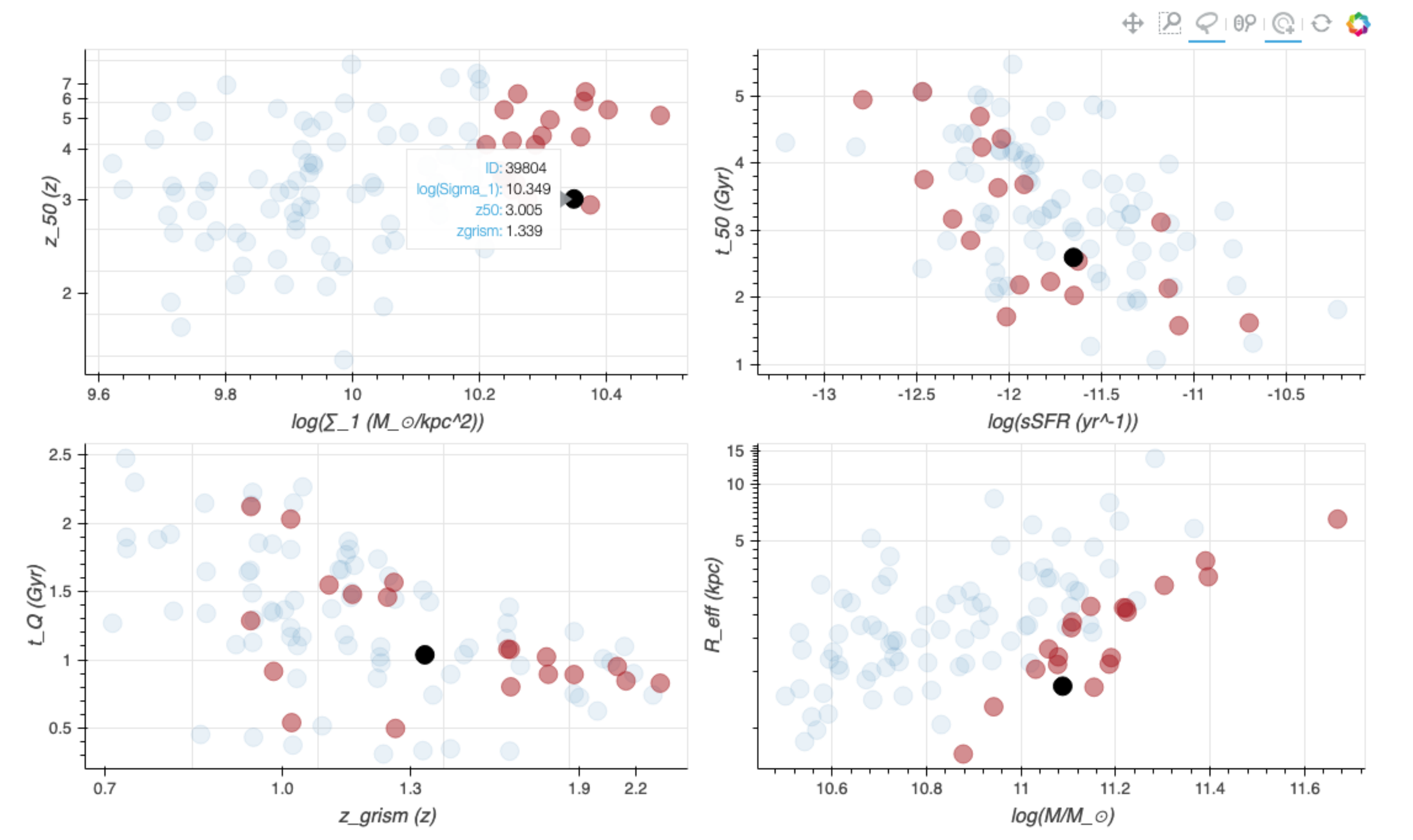}
\caption{Here we show an example of the usage of our interactive appendix (all data shown here were discussed in the text). Using the lasso tool we
select all galaxies with \Lsig\ $> 10.2$, this population is highlighted in all plots. Additionally, by hovering over a galaxy, we get more information about it. \label{appendix_main}}
\end{figure}

\begin{figure}
\epsscale{0.65}
\plotone{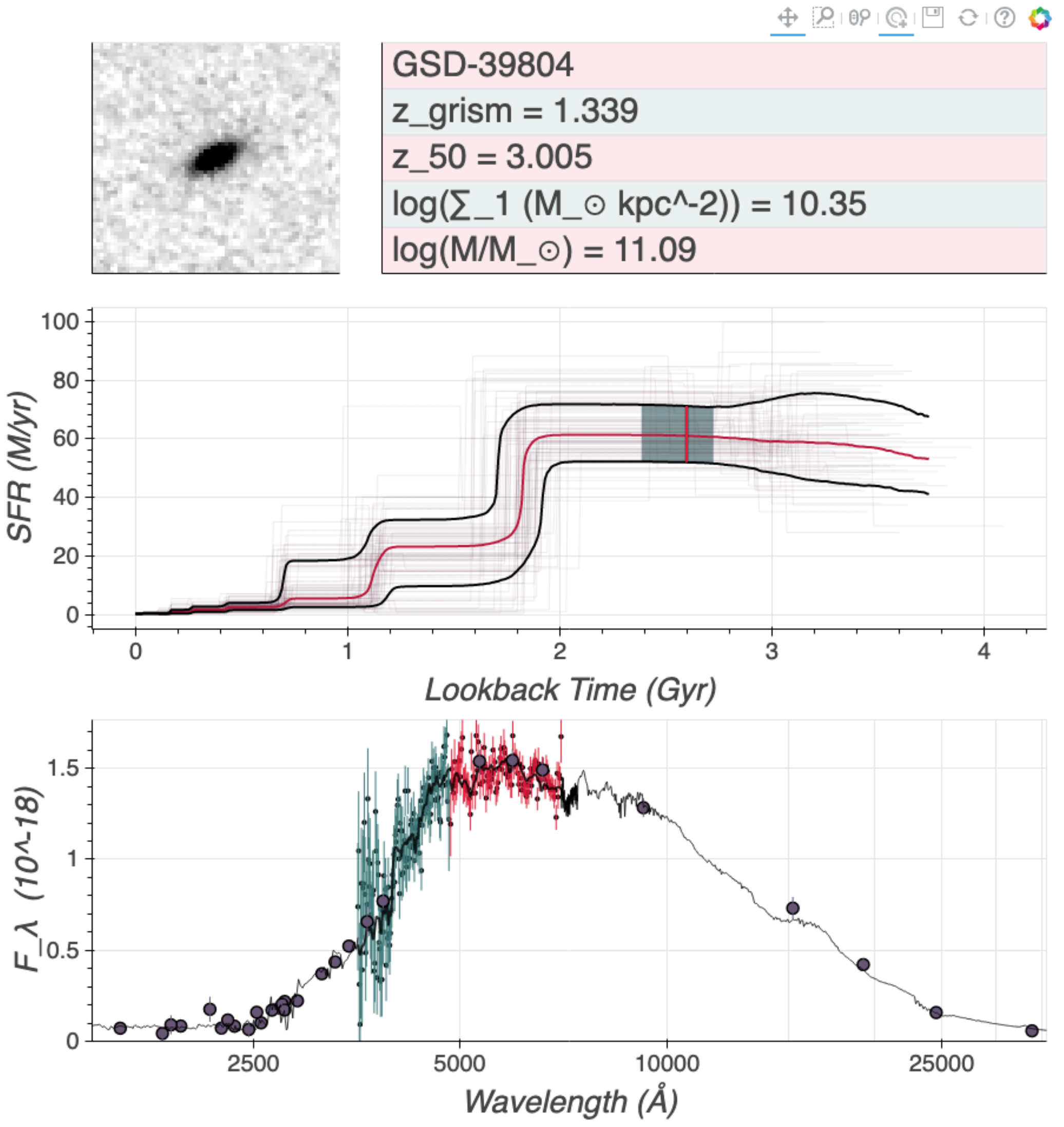}
\caption{Example bio page for galaxy GSD-39804. When a point in Figure \ref{appendix_main} is clicked it will bring up the galaxies bio page. These bio pages includes the galaxy's morphology, a data table, interactive SFH, and interactive spectra with best fit model.\label{appendix_bio}}
\end{figure}

\begin{figure}
\epsscale{0.65}
\plotone{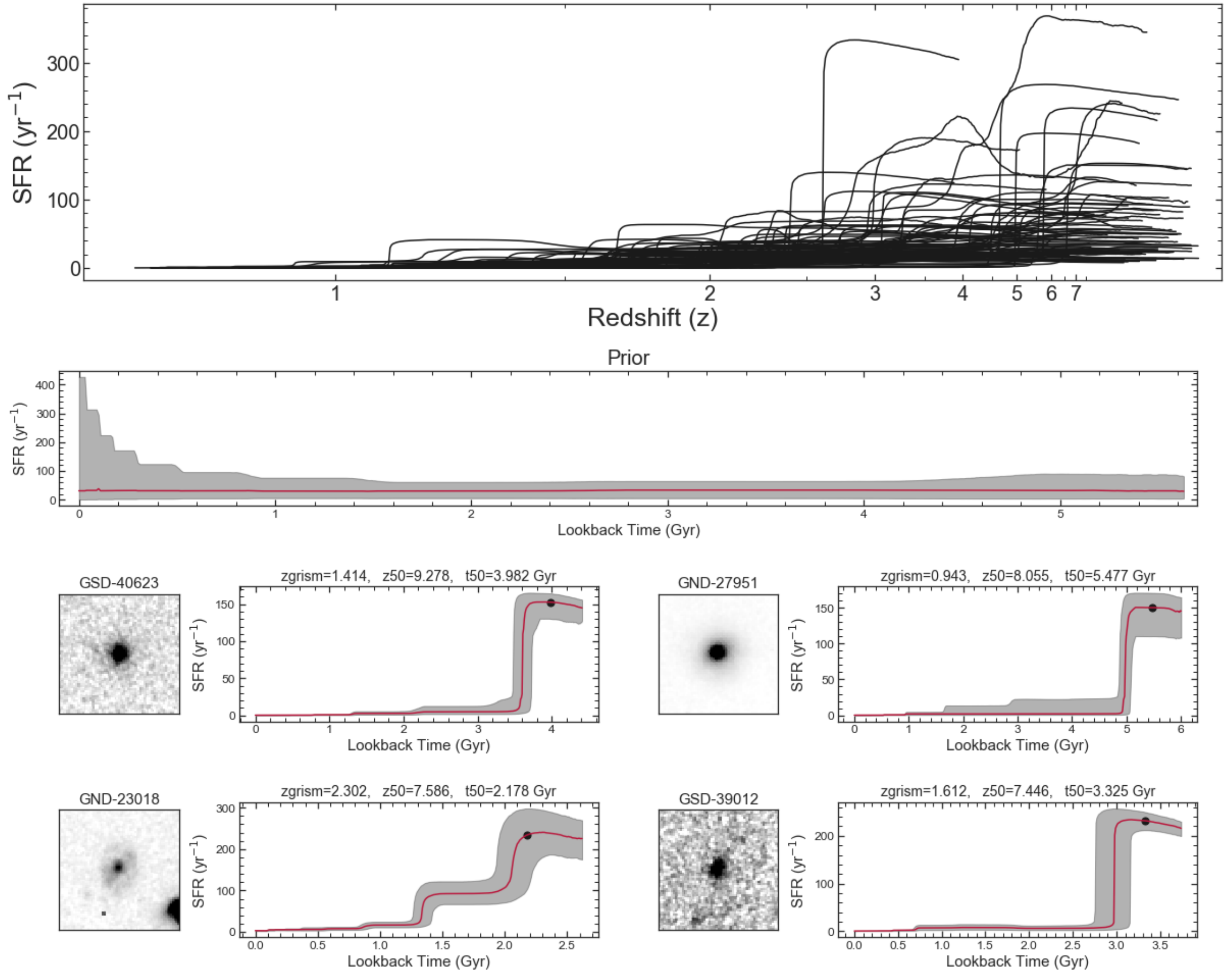}
\caption{A representative plot for a figure included in our appendix. The top plot shows all star-formation histories, plotted at their appropriate redshifts. The next plot down shows the prior we used to fit our "non-parametric" star-formation histories. The following plots are then the galaxy cutouts and star-formation histories for each galaxy, with their formation redshift marked with a point, and relevant information shown at the top of the figure (ordered by \zfif). \label{appendix_sfh} }
\end{figure}
\appendix

\section{Data Tables}\label{section_appendix_tables}
\editone{We report here the catalog for the galaxies in the sample
  used here.  These include two tables.   Table~\ref{table1} reports
  values for each galaxy derived without the analysis of the grism
  data  (including galaxy ID numbers, coordinates, photometric
  redshift, photometric masses, and circularized radii).
  Table~\ref{table2}  reports values derived from our fits to the
  photometry and grism data here (including redshifts, masses,
  specific star-formation rates, dust attenuations, stellar-mass
  surface densities, quenching timescales, and formation redshifts).}

\section{Interactive Online Model Fits for the Galaxy Sample}\label{sec_appendix_online}
\editone{We include with this paper an interactive appendix, which
  shows the properties and model fits for all the galaxies in
  our sample.  The appendix is available here: 
\href{https://vince-ec.github.io/appendix/appendix}{interactive online
  appendix}\footnote{Also available at \href{https://vince-ec.github.io/appendix/appendix}{https://vince-ec.github.io/appendix/appendix}}.  At this link the
reader can see galaxy properties on multiple plots simultaneously (Figure \ref{appendix_main}), and also
access individual galaxy morphology, photometric and spectroscopic
data and model fits (Figure \ref{appendix_bio}).  The user can interact with the star-formation history, 
and spectral energy distribution.   The online material also includes
a hyperlink to show \href{https://vince-ec.github.io/appendix/fullfig}{all galaxy SFHs
and morphologies ordered by $z_{50}$}\footnote{Also available at \href{https://vince-ec.github.io/appendix/fullfig}{https://vince-ec.github.io/appendix/fullfig}} on a single figure (Figure \ref{appendix_sfh}).}

\startlongtable
  \begin{deluxetable*}{lccccc}
\tabletypesize{\small}
\tablecaption{Catalog Properties of Quiescent Galaxy Sample\label{table1}}
\tablecolumns{6}
\tablewidth{0pt}
\tablehead{
\colhead{ID}  &
\colhead{RA}  &
\colhead{DEC } &
\colhead{$z_{\mathrm{phot}}$} &
\colhead{$\log(M_{\mathrm{phot}})$} &
\colhead{$R_{\mathrm{eff}}$}  \\
\colhead{} & 
\colhead{(deg, J2000)} &
\colhead{(deg, J2000)} &
\colhead{} &
\colhead{$(\log M_\odot)$} &
\colhead{(kpc)} \\
\colhead{(1)} & 
\colhead{(2)} & 
\colhead{(3)} &
\colhead{(4)} &
\colhead{(5)} &
\colhead{(6)} }
\startdata
GND-29879 & 189.254227 & 62.291579 & $0.69_{0.01}^{0.01}$ & 10.9 & $1.12_{0.01}^{0.13}$ \\
GSD-41147 & 53.081634 & -27.717718 & $0.70_{0.01}^{0.01}$ & 10.8 & $1.46_{0.01}^{0.09}$ \\
GSD-47140 & 53.131853 & -27.687304 & $0.73_{0.01}^{0.01}$ & 10.8 & $2.91_{0.04}^{0.22}$ \\
GSD-46001 & 53.120312 & -27.691486 & $0.72_{0.01}^{0.01}$ & 11.2 & $3.00_{0.02}^{0.19}$ \\
GND-27006 & 189.263714 & 62.275807 & $0.71_{0.02}^{0.02}$ & 10.8 & $1.11_{0.01}^{0.11}$ \\
GND-22358 & 189.081040 & 62.251545 & $0.82_{0.02}^{0.02}$ & 10.7 & $1.66_{0.02}^{0.12}$ \\
GND-36838 & 189.251622 & 62.344526 & $0.80_{0.02}^{0.01}$ & 10.8 & $1.46_{0.03}^{0.12}$ \\
GND-37186 & 189.243199 & 62.349892 & $0.80_{0.01}^{0.02}$ & 11.0 & $1.68_{0.03}^{0.15}$ \\
GND-13774 & 189.179829 & 62.211733 & $0.83_{0.01}^{0.01}$ & 11.1 & $1.78_{0.01}^{0.14}$ \\
GND-32108 & 189.277164 & 62.305097 & $0.82_{0.01}^{0.01}$ & 10.7 & $1.77_{0.03}^{0.15}$ \\
GND-23459 & 189.310355 & 62.258286 & $0.86_{0.01}^{0.01}$ & 11.0 & $1.99_{0.02}^{0.16}$ \\
GND-24795 & 189.202555 & 62.264622 & $0.85_{0.01}^{0.01}$ & 10.9 & $2.31_{0.01}^{0.14}$ \\
GND-14158 & 189.192218 & 62.212927 & $0.87_{0.01}^{0.01}$ & 10.6 & $2.91_{0.04}^{0.21}$ \\
GND-29183 & 189.245464 & 62.287267 & $0.94_{0.03}^{0.02}$ & 10.6 & $0.74_{0.01}^{0.06}$ \\
GND-24177 & 189.343191 & 62.262053 & $0.91_{0.02}^{0.01}$ & 11.0 & $4.70_{0.06}^{0.33}$ \\
GND-23081 & 189.334875 & 62.255800 & $0.90_{0.02}^{0.02}$ & 11.2 & $2.23_{0.02}^{0.17}$ \\
GND-22213 & 189.201365 & 62.252076 & $0.88_{0.01}^{0.01}$ & 11.2 & $1.10_{0.01}^{0.11}$ \\
GND-33453 & 189.264307 & 62.314325 & $0.88_{0.02}^{0.03}$ & 10.7 & $2.45_{0.05}^{0.17}$ \\
GND-23758 & 189.217983 & 62.260352 & $0.89_{0.02}^{0.02}$ & 11.2 & $7.96_{0.07}^{0.54}$ \\
GND-22246 & 189.220896 & 62.252424 & $0.87_{0.01}^{0.01}$ & 11.1 & $3.58_{0.03}^{0.22}$ \\
GND-26673 & 189.279210 & 62.274483 & $0.92_{0.02}^{0.02}$ & 10.8 & $3.22_{0.05}^{0.24}$ \\
GND-27951 & 189.226202 & 62.281984 & $0.94_{0.01}^{0.02}$ & 11.1 & $3.54_{0.03}^{0.23}$ \\
GND-37340 & 189.289100 & 62.352859 & $0.88_{0.01}^{0.01}$ & 11.0 & $1.75_{0.03}^{0.12}$ \\
GND-12793 & 189.236022 & 62.205604 & $0.96_{0.02}^{0.01}$ & 10.9 & $2.56_{0.03}^{0.17}$ \\
GSD-38191 & 53.141108 & -27.732652 & $0.97_{0.01}^{0.01}$ & 10.7 & $1.27_{0.02}^{0.08}$ \\
GSD-39850 & 53.173100 & -27.724355 & $0.96_{0.01}^{0.01}$ & 10.7 & $0.91_{0.01}^{0.08}$ \\
GND-36161 & 189.201410 & 62.336077 & $0.92_{0.02}^{0.02}$ & 10.7 & $1.20_{0.03}^{0.08}$ \\
GSD-19148 & 53.164983 & -27.819326 & $0.97_{0.01}^{0.01}$ & 11.5 & $3.90_{0.02}^{0.27}$ \\
GND-22363 & 189.169647 & 62.251296 & $0.99_{0.02}^{0.01}$ & 10.7 & $1.00_{0.01}^{0.07}$ \\
GND-27185 & 189.242059 & 62.277510 & $1.04_{0.01}^{0.02}$ & 11.2 & $1.72_{0.02}^{0.16}$ \\
GSD-42221 & 53.079234 & -27.711869 & $1.03_{0.02}^{0.02}$ & 10.8 & $5.14_{0.10}^{0.39}$ \\
GND-16758 & 189.162357 & 62.224840 & $0.98_{0.01}^{0.01}$ & 11.1 & $1.45_{0.01}^{0.19}$ \\
GND-12078 & 189.166744 & 62.202054 & $0.99_{0.01}^{0.01}$ & 10.7 & $1.99_{0.02}^{0.13}$ \\
GSD-39170 & 53.041826 & -27.725868 & $1.03_{0.01}^{0.01}$ & 11.4 & $3.20_{0.02}^{0.21}$ \\
GSD-43615 & 53.093057 & -27.707368 & $1.03_{0.01}^{0.01}$ & 10.9 & $1.50_{0.02}^{0.13}$ \\
GND-22633 & 189.161700 & 62.252923 & $1.00_{0.02}^{0.03}$ & 10.7 & $1.40_{0.02}^{0.16}$ \\
GSD-39241 & 53.042327 & -27.726209 & $1.03_{0.01}^{0.01}$ & 11.1 & $2.23_{0.02}^{0.14}$ \\
GSD-39631 & 53.042169 & -27.725928 & $0.99_{0.01}^{0.02}$ & 11.0 & $2.24_{0.03}^{0.18}$ \\
GND-37955 & 189.337824 & 62.371137 & $0.98_{0.01}^{0.04}$ & 11.0 & $3.17_{0.07}^{0.22}$ \\
GND-37210 & 189.252761 & 62.350806 & $1.04_{0.01}^{0.02}$ & 11.2 & $2.67_{0.04}^{0.19}$ \\
GSD-45972 & 53.115984 & -27.693568 & $1.03_{0.01}^{0.01}$ & 11.1 & $8.36_{0.13}^{0.65}$ \\
GSD-44620 & 53.249645 & -27.702048 & $1.09_{0.02}^{0.01}$ & 10.7 & $4.11_{0.05}^{0.31}$ \\
GSD-29928 & 53.154965 & -27.768904 & $1.09_{0.01}^{0.01}$ & 11.7 & $6.50_{0.01}^{0.47}$ \\
GND-30358 & 189.299204 & 62.293310 & $0.94_{0.03}^{0.02}$ & 10.6 & $1.17_{0.02}^{0.08}$ \\
GND-23857 & 189.070894 & 62.259299 & $1.15_{0.04}^{0.04}$ & 10.6 & $1.62_{0.04}^{0.11}$ \\
GSD-47691 & 53.273156 & -27.681599 & $1.12_{0.01}^{0.01}$ & 11.2 & $5.23_{0.02}^{0.35}$ \\
GND-21724 & 189.063257 & 62.248675 & $1.10_{0.02}^{0.02}$ & 11.0 & $2.69_{0.05}^{0.20}$ \\
GND-37325 & 189.251371 & 62.351582 & $1.19_{0.04}^{0.03}$ & 10.6 & $0.81_{0.03}^{0.07}$ \\
GND-22027 & 189.065790 & 62.249816 & $1.14_{0.03}^{0.02}$ & 10.9 & $1.68_{0.03}^{0.11}$ \\
GND-34694 & 189.147840 & 62.323647 & $1.07_{0.02}^{0.02}$ & 11.1 & $4.62_{0.05}^{0.36}$ \\
GND-38102 & 189.339219 & 62.375874 & $1.24_{0.02}^{0.01}$ & 10.7 & $0.58_{0.03}^{0.06}$ \\
GND-28451 & 189.247715 & 62.282931 & $1.14_{0.01}^{0.02}$ & 10.7 & $1.01_{0.01}^{0.07}$ \\
GND-20432 & 189.362767 & 62.242309 & $1.14_{0.02}^{0.02}$ & 11.1 & $1.15_{0.01}^{0.07}$ \\
GND-17746 & 189.049436 & 62.228979 & $1.16_{0.01}^{0.02}$ & 11.1 & $1.32_{0.04}^{0.16}$ \\
GSD-39805 & 53.163237 & -27.724724 & $1.15_{0.01}^{0.02}$ & 10.8 & $2.33_{0.06}^{0.16}$ \\
GSD-40476 & 53.108262 & -27.721924 & $1.18_{0.01}^{0.02}$ & 10.9 & $1.13_{0.02}^{0.10}$ \\
GSD-37828 & 53.158121 & -27.734502 & $1.20_{0.02}^{0.02}$ & 10.7 & $1.09_{0.03}^{0.08}$ \\
GSD-40597 & 53.148451 & -27.719472 & $1.23_{0.01}^{0.01}$ & 11.4 & $1.79_{0.02}^{0.14}$ \\
GND-34419 & 189.311828 & 62.320264 & $1.20_{0.01}^{0.01}$ & 10.8 & $0.71_{0.02}^{0.05}$ \\
GND-13191 & 189.217041 & 62.207326 & $1.28_{0.02}^{0.02}$ & 10.8 & $1.77_{0.04}^{0.14}$ \\
GND-14713 & 189.236333 & 62.214608 & $1.23_{0.02}^{0.03}$ & 10.8 & $1.51_{0.03}^{0.11}$ \\
GND-17070 & 189.268086 & 62.226445 & $1.24_{0.02}^{0.01}$ & 11.3 & $0.83_{0.01}^{0.10}$ \\
GSD-38785 & 53.168249 & -27.727300 & $1.14_{0.04}^{0.04}$ & 11.1 & $3.43_{0.04}^{0.26}$ \\
GND-21156 & 189.239409 & 62.247548 & $1.21_{0.02}^{0.02}$ & 11.4 & $2.88_{0.03}^{0.24}$ \\
GSD-35774 & 53.158775 & -27.742385 & $1.23_{0.01}^{0.01}$ & 11.3 & $6.35_{0.07}^{0.40}$ \\
GND-37686 & 189.274474 & 62.360820 & $1.28_{0.02}^{0.02}$ & 11.2 & $1.84_{0.05}^{0.14}$ \\
GSD-40862 & 53.048020 & -27.719743 & $1.34_{0.01}^{0.02}$ & 11.2 & $2.71_{0.03}^{0.21}$ \\
GSD-46066 & 53.061039 & -27.693501 & $1.32_{0.01}^{0.02}$ & 11.2 & $1.74_{0.03}^{0.11}$ \\
GSD-39804 & 53.178423 & -27.724640 & $1.36_{0.01}^{0.01}$ & 11.2 & $0.84_{0.01}^{0.11}$ \\
GSD-45775 & 53.158558 & -27.694968 & $1.37_{0.01}^{0.02}$ & 11.4 & $13.68_{0.21}^{0.93}$ \\
GND-36530 & 189.275620 & 62.340723 & $1.39_{0.02}^{0.03}$ & 11.1 & $6.06_{0.24}^{0.49}$ \\
GSD-40623 & 53.130480 & -27.721152 & $1.43_{0.02}^{0.02}$ & 11.1 & $2.21_{0.04}^{0.15}$ \\
GND-24345 & 189.244758 & 62.261225 & $1.35_{0.03}^{0.03}$ & 10.6 & $0.49_{0.02}^{0.04}$ \\
GND-16574 & 189.233886 & 62.223678 & $1.50_{0.03}^{0.03}$ & 10.7 & $0.77_{0.03}^{0.06}$ \\
GND-21427 & 189.368121 & 62.247344 & $1.50_{0.02}^{0.02}$ & 11.0 & $2.34_{0.05}^{0.15}$ \\
GSD-40223 & 53.124956 & -27.722957 & $1.66_{0.03}^{0.02}$ & 11.0 & $1.06_{0.02}^{0.08}$ \\
GSD-39649 & 53.059630 & -27.725792 & $1.66_{0.02}^{0.01}$ & 10.9 & $0.74_{0.02}^{0.05}$ \\
GSD-42487 & 53.116396 & -27.712701 & $1.69_{0.03}^{0.02}$ & 11.0 & $0.65_{0.01}^{0.05}$ \\
GSD-38843 & 53.107039 & -27.729749 & $1.61_{0.03}^{0.04}$ & 10.6 & $1.31_{0.06}^{0.13}$ \\
GSD-39012 & 53.064240 & -27.727621 & $1.62_{0.04}^{0.04}$ & 11.3 & $1.62_{0.05}^{0.15}$ \\
GSD-41520 & 53.152726 & -27.716251 & $1.64_{0.02}^{0.02}$ & 11.2 & $1.20_{0.02}^{0.10}$ \\
GSD-44042 & 53.104570 & -27.705421 & $1.81_{0.02}^{0.02}$ & 11.4 & $2.19_{0.03}^{0.15}$ \\
GND-33775 & 189.188648 & 62.315319 & $1.65_{0.08}^{0.06}$ & 10.7 & $0.60_{0.02}^{0.05}$ \\
GSD-42615 & 53.127414 & -27.712062 & $1.67_{0.03}^{0.03}$ & 11.2 & $1.03_{0.02}^{0.07}$ \\
GSD-41148 & 53.127925 & -27.718885 & $1.79_{0.02}^{0.02}$ & 11.4 & $2.19_{0.03}^{0.14}$ \\
GND-33780 & 189.202025 & 62.317153 & $1.92_{0.05}^{0.06}$ & 11.6 & $5.79_{0.09}^{0.37}$ \\
GND-17735 & 189.060905 & 62.228977 & $1.90_{0.03}^{0.02}$ & 11.1 & $1.09_{0.01}^{0.08}$ \\
GND-19850 & 189.090085 & 62.239244 & $1.85_{0.02}^{0.02}$ & 10.9 & $0.80_{0.01}^{0.07}$ \\
GSD-24569 & 53.158798 & -27.797153 & $1.90_{0.02}^{0.02}$ & 11.0 & $0.52_{0.01}^{0.04}$ \\
GSD-24315 & 53.162991 & -27.797654 & $2.01_{0.02}^{0.02}$ & 10.7 & $0.42_{0.01}^{0.04}$ \\
GND-14132 & 189.190249 & 62.211662 & $2.01_{0.03}^{0.03}$ & 11.1 & $1.11_{0.02}^{0.09}$ \\
GSD-43572 & 53.142153 & -27.707427 & $2.05_{0.04}^{0.03}$ & 11.2 & $3.14_{0.12}^{0.26}$ \\
GND-21738 & 189.210937 & 62.248818 & $2.11_{0.02}^{0.03}$ & 11.4 & $1.19_{0.02}^{0.10}$ \\
GND-32933 & 189.156358 & 62.309106 & $2.13_{0.04}^{0.04}$ & 10.7 & $1.06_{0.04}^{0.10}$ \\
GND-17599 & 189.121464 & 62.228903 & $2.12_{0.02}^{0.01}$ & 11.0 & $0.36_{0.01}^{0.04}$ \\
GSD-44133 & 53.110407 & -27.703706 & $2.09_{0.01}^{0.02}$ & 10.4 & $1.59_{0.03}^{0.13}$ \\
GND-23018 & 189.277544 & 62.254617 & $2.25_{0.03}^{0.03}$ & 11.3 & $2.39_{0.04}^{0.17}$ \\
GSD-48464 & 53.144819 & -27.682470 & $2.34_{0.03}^{0.03}$ & 11.4 & $2.08_{0.06}^{0.18}$ \\
\enddata
\tablecomments{\editone{(1) catalog ID number (matching those in \citet{skel14}); 
(2) right ascension; (3) declination; (4) photometric redshift; 
(5) stellar mass from Eazy-py; (6) \textit{circularized} effective radius (derived from \cite{vanw14} 
and defined as r$\sqrt{b/a}$, where $r$ is the radius of the semi-major axis in kpc, $b/a$ 
is the axis ratio)}}
\end{deluxetable*}


\startlongtable
\begin{deluxetable*}{lcccccccc}
\tabletypesize{\small}
\tablecaption{Catalog Properties of Quiescent Galaxy Sample\label{table1}}
\tablecolumns{9}
\tablewidth{0pt}
\tablecaption{Derived Properties of Quiescent Galaxy Sample\label{table2}}
\tablehead{
\colhead{ID} &
\colhead{$z_\mathrm{grism}$} &
\colhead{$\log(M_\mathrm{grism})$} &
\colhead{$\log$  sSFR} &
\colhead{$A_V$} &
\colhead{$\log(\Sigma_1)$} &
\colhead{$t_{Q}$} &
\colhead{$z_{50}$} \\
\colhead{} &
\colhead{} &
\colhead{$(\log M_\odot)$} &
\colhead{$(\log \mathrm{yr^{-1}} )$} &
\colhead{(mag)} &
\colhead{$(\log M_\odot\ \mathrm{kpc^{-2}} )$} &
\colhead{(Gyr)} &
\colhead{}  \\
\colhead{(1)} & 
\colhead{(2)} & 
\colhead{(3)} &
\colhead{(4)} &
\colhead{(5)} &
\colhead{(6)} &
\colhead{(7)} &
\colhead{(8)} }
\startdata
GND-29879 & $0.711_{0.002}^{0.001}$ & $10.80_{0.03}^{0.04}$ & $-12.2_{0.5}^{0.4}$ & $0.22_{0.14}^{0.30}$ & $9.99_{0.05}^{0.05}$ & $1.3_{0.8}^{0.4}$ & $1.6_{0.2}^{0.2}$ \\
GSD-41147 & $0.730_{0.002}^{0.002}$ & $10.74_{0.02}^{0.02}$ & $-11.6_{0.1}^{0.4}$ & $0.20_{0.15}^{0.24}$ & $9.96_{0.02}^{0.02}$ & $2.5_{0.5}^{0.3}$ & $2.1_{0.3}^{0.2}$ \\
GSD-47140 & $0.731_{0.002}^{0.002}$ & $10.70_{0.02}^{0.03}$ & $-12.0_{0.1}^{0.2}$ & $0.00_{0.00}^{0.02}$ & $9.71_{0.04}^{0.04}$ & $1.9_{1.1}^{0.4}$ & $2.8_{0.8}^{0.2}$ \\
GSD-46001 & $0.732_{0.001}^{0.001}$ & $11.10_{0.02}^{0.03}$ & $-11.7_{0.1}^{0.5}$ & $0.43_{0.35}^{0.47}$ & $10.05_{0.03}^{0.03}$ & $1.8_{0.8}^{0.3}$ & $1.9_{0.4}^{0.1}$ \\
GND-27006 & $0.743_{0.001}^{0.001}$ & $10.88_{0.02}^{0.04}$ & $-12.0_{0.2}^{0.6}$ & $0.26_{0.20}^{0.34}$ & $10.13_{0.04}^{0.04}$ & $2.3_{1.0}^{0.5}$ & $3.0_{0.8}^{0.5}$ \\
GND-22358 & $0.779_{0.005}^{0.005}$ & $10.70_{0.03}^{0.03}$ & $-11.6_{0.2}^{0.1}$ & $0.18_{0.13}^{0.26}$ & $9.77_{0.04}^{0.04}$ & $1.9_{0.6}^{0.5}$ & $3.1_{0.6}^{0.5}$ \\
GND-36838 & $0.799_{0.002}^{0.002}$ & $10.73_{0.03}^{0.04}$ & $-12.4_{0.3}^{0.6}$ & $0.03_{0.01}^{0.10}$ & $9.94_{0.04}^{0.04}$ & $1.9_{0.8}^{0.5}$ & $3.6_{1.0}^{0.6}$ \\
GND-37186 & $0.804_{0.001}^{0.001}$ & $10.91_{0.02}^{0.02}$ & $-11.9_{0.1}^{0.6}$ & $0.01_{0.01}^{0.06}$ & $10.05_{0.03}^{0.03}$ & $1.4_{0.5}^{0.4}$ & $2.4_{0.3}^{0.2}$ \\
GND-13774 & $0.849_{0.001}^{0.001}$ & $11.02_{0.03}^{0.02}$ & $-12.4_{0.1}^{1.1}$ & $0.01_{0.01}^{0.01}$ & $10.14_{0.03}^{0.03}$ & $0.5_{0.6}^{0.4}$ & $2.8_{0.4}^{0.2}$ \\
GND-32108 & $0.855_{0.002}^{0.004}$ & $10.66_{0.03}^{0.03}$ & $-11.2_{0.2}^{0.1}$ & $0.33_{0.26}^{0.51}$ & $9.72_{0.04}^{0.04}$ & $2.1_{1.1}^{0.4}$ & $2.6_{0.7}^{0.4}$ \\
GND-23459 & $0.858_{0.001}^{0.001}$ & $10.93_{0.02}^{0.02}$ & $-12.2_{0.4}^{0.5}$ & $0.41_{0.31}^{0.48}$ & $10.01_{0.03}^{0.03}$ & $1.6_{0.8}^{0.4}$ & $3.1_{0.7}^{0.3}$ \\
GND-24795 & $0.858_{0.002}^{0.003}$ & $10.84_{0.03}^{0.03}$ & $-11.9_{0.3}^{0.5}$ & $0.00_{0.00}^{0.07}$ & $9.73_{0.05}^{0.05}$ & $1.3_{0.8}^{0.3}$ & $1.8_{0.3}^{0.1}$ \\
GND-14158 & $0.911_{0.004}^{0.005}$ & $10.58_{0.02}^{0.02}$ & $-11.3_{0.1}^{0.3}$ & $0.00_{0.00}^{0.02}$ & $9.71_{0.03}^{0.03}$ & $1.1_{1.0}^{0.2}$ & $1.9_{0.4}^{0.1}$ \\
GND-29183 & $0.933_{0.002}^{0.007}$ & $10.50_{0.03}^{0.03}$ & $-11.5_{0.6}^{0.3}$ & $0.45_{0.34}^{0.58}$ & $9.83_{0.03}^{0.03}$ & $1.6_{0.9}^{0.8}$ & $2.2_{0.1}^{0.7}$ \\
GND-24177 & $0.937_{0.002}^{0.002}$ & $10.96_{0.03}^{0.03}$ & $-12.2_{0.2}^{0.5}$ & $0.00_{0.00}^{0.04}$ & $9.93_{0.03}^{0.03}$ & $1.7_{0.7}^{0.4}$ & $3.7_{0.8}^{0.5}$ \\
GND-23081 & $0.938_{0.001}^{0.001}$ & $11.15_{0.02}^{0.02}$ & $-12.2_{0.2}^{0.2}$ & $0.17_{0.14}^{0.22}$ & $10.25_{0.03}^{0.03}$ & $1.3_{1.8}^{0.5}$ & $4.2_{0.7}^{1.0}$ \\
GND-22213 & $0.938_{0.001}^{0.001}$ & $11.19_{0.02}^{0.02}$ & $-12.2_{0.1}^{0.5}$ & $0.01_{0.01}^{0.09}$ & $10.40_{0.03}^{0.03}$ & $2.1_{0.5}^{0.4}$ & $5.4_{1.1}^{1.2}$ \\
GND-33453 & $0.939_{0.003}^{0.003}$ & $10.62_{0.05}^{0.04}$ & $-12.2_{0.6}^{0.3}$ & $0.18_{0.12}^{0.26}$ & $9.72_{0.05}^{0.05}$ & $2.1_{1.2}^{1.0}$ & $3.2_{0.9}^{2.2}$ \\
GND-23758 & $0.941_{0.002}^{0.002}$ & $11.19_{0.04}^{0.03}$ & $-11.9_{0.2}^{0.4}$ & $0.49_{0.35}^{0.55}$ & $10.13_{0.03}^{0.03}$ & $1.5_{0.6}^{0.3}$ & $3.3_{0.6}^{0.4}$ \\
GND-22246 & $0.942_{0.001}^{0.001}$ & $11.05_{0.02}^{0.01}$ & $-12.2_{0.3}^{0.2}$ & $0.00_{0.00}^{0.02}$ & $10.04_{0.02}^{0.02}$ & $2.2_{0.9}^{0.6}$ & $5.2_{1.1}^{1.7}$ \\
GND-26673 & $0.942_{0.001}^{0.001}$ & $10.72_{0.02}^{0.01}$ & $-11.9_{0.5}^{0.1}$ & $0.07_{0.04}^{0.10}$ & $9.82_{0.02}^{0.02}$ & $1.1_{0.4}^{0.2}$ & $2.1_{0.2}^{0.1}$ \\
GND-27951 & $0.943_{0.004}^{0.004}$ & $11.19_{0.01}^{0.03}$ & $-12.3_{0.5}^{0.1}$ & $0.09_{0.06}^{0.13}$ & $10.20_{0.03}^{0.03}$ & $0.4_{0.4}^{0.2}$ & $8.1_{1.4}^{1.6}$ \\
GND-37340 & $0.945_{0.001}^{0.001}$ & $11.01_{0.01}^{0.01}$ & $-12.4_{0.3}^{0.1}$ & $0.02_{0.01}^{0.02}$ & $10.17_{0.02}^{0.02}$ & $1.3_{0.2}^{0.2}$ & $3.7_{0.3}^{0.2}$ \\
GND-12793 & $0.953_{0.002}^{0.004}$ & $10.86_{0.03}^{0.02}$ & $-11.4_{0.2}^{0.2}$ & $0.14_{0.08}^{0.23}$ & $9.91_{0.03}^{0.03}$ & $1.9_{0.4}^{0.2}$ & $2.7_{0.3}^{0.2}$ \\
GSD-38191 & $0.977_{0.002}^{0.003}$ & $10.60_{0.03}^{0.02}$ & $-11.6_{0.1}^{0.7}$ & $0.21_{0.14}^{0.24}$ & $9.76_{0.03}^{0.03}$ & $1.4_{0.6}^{0.2}$ & $2.8_{0.3}^{0.2}$ \\
GSD-39850 & $0.980_{0.001}^{0.001}$ & $10.67_{0.02}^{0.02}$ & $-12.3_{0.4}^{0.4}$ & $0.04_{0.02}^{0.08}$ & $9.97_{0.03}^{0.03}$ & $1.8_{0.6}^{0.2}$ & $2.3_{0.3}^{0.1}$ \\
GND-36161 & $0.981_{0.034}^{0.010}$ & $10.73_{0.04}^{0.04}$ & $-11.7_{0.3}^{0.5}$ & $0.00_{0.00}^{0.04}$ & $9.91_{0.05}^{0.05}$ & $1.3_{1.1}^{0.5}$ & $2.6_{0.7}^{0.4}$ \\
GSD-19148 & $0.982_{0.001}^{0.001}$ & $11.39_{0.02}^{0.02}$ & $-12.6_{0.4}^{0.2}$ & $0.18_{0.11}^{0.24}$ & $10.38_{0.03}^{0.03}$ & $0.9_{1.2}^{0.5}$ & $2.9_{0.3}^{0.2}$ \\
GND-22363 & $1.004_{0.002}^{0.002}$ & $10.68_{0.01}^{0.01}$ & $-11.9_{0.1}^{0.2}$ & $0.01_{0.01}^{0.02}$ & $9.94_{0.02}^{0.02}$ & $1.4_{0.4}^{0.2}$ & $3.7_{0.5}^{0.1}$ \\
GND-27185 & $1.016_{0.002}^{0.003}$ & $11.11_{0.03}^{0.03}$ & $-11.8_{0.2}^{0.5}$ & $0.35_{0.26}^{0.43}$ & $10.28_{0.04}^{0.04}$ & $2.0_{0.5}^{0.4}$ & $3.0_{0.5}^{0.5}$ \\
GSD-42221 & $1.016_{0.004}^{0.003}$ & $10.68_{0.04}^{0.02}$ & $-11.6_{0.2}^{0.1}$ & $0.02_{0.01}^{0.09}$ & $9.72_{0.03}^{0.03}$ & $1.2_{0.4}^{0.4}$ & $3.1_{0.4}^{0.3}$ \\
GND-16758 & $1.016_{0.001}^{0.001}$ & $10.98_{0.04}^{0.03}$ & $-11.7_{0.4}^{0.2}$ & $0.56_{0.51}^{0.62}$ & $10.07_{0.06}^{0.06}$ & $1.2_{0.6}^{0.5}$ & $2.5_{0.4}^{0.4}$ \\
GND-12078 & $1.016_{0.001}^{0.002}$ & $10.80_{0.03}^{0.02}$ & $-10.9_{0.1}^{0.9}$ & $0.27_{0.17}^{0.36}$ & $9.89_{0.03}^{0.03}$ & $1.8_{0.5}^{0.2}$ & $2.1_{0.3}^{0.2}$ \\
GSD-39170 & $1.018_{0.001}^{0.001}$ & $11.40_{0.02}^{0.02}$ & $-12.5_{0.1}^{0.6}$ & $0.02_{0.01}^{0.05}$ & $10.37_{0.03}^{0.03}$ & $0.5_{1.6}^{0.2}$ & $6.4_{1.4}^{1.5}$ \\
GSD-43615 & $1.021_{0.002}^{0.001}$ & $10.88_{0.01}^{0.02}$ & $-11.8_{0.4}^{0.1}$ & $0.39_{0.34}^{0.42}$ & $9.99_{0.04}^{0.04}$ & $0.4_{0.3}^{0.3}$ & $5.7_{0.8}^{0.6}$ \\
GND-22633 & $1.022_{0.007}^{0.006}$ & $10.72_{0.08}^{0.03}$ & $-10.8_{0.3}^{0.4}$ & $0.87_{0.73}^{0.93}$ & $9.79_{0.08}^{0.08}$ & $2.1_{0.9}^{0.8}$ & $2.6_{0.7}^{0.8}$ \\
GSD-39241 & $1.024_{0.001}^{0.003}$ & $11.11_{0.02}^{0.02}$ & $-11.9_{0.1}^{0.3}$ & $0.26_{0.22}^{0.30}$ & $10.15_{0.03}^{0.03}$ & $1.4_{0.4}^{0.2}$ & $3.9_{0.7}^{0.3}$ \\
GSD-39631 & $1.029_{0.003}^{0.003}$ & $10.90_{0.03}^{0.02}$ & $-12.9_{0.6}^{0.7}$ & $0.25_{0.22}^{0.36}$ & $9.98_{0.03}^{0.03}$ & $0.9_{1.8}^{0.4}$ & $4.2_{1.1}^{0.7}$ \\
GND-37955 & $1.030_{0.004}^{0.004}$ & $11.06_{0.04}^{0.04}$ & $-12.0_{0.5}^{0.2}$ & $0.23_{0.17}^{0.39}$ & $9.95_{0.05}^{0.05}$ & $1.1_{1.6}^{0.3}$ & $4.9_{1.2}^{0.9}$ \\
GND-37210 & $1.040_{0.002}^{0.002}$ & $11.12_{0.02}^{0.02}$ & $-12.2_{0.2}^{0.3}$ & $0.00_{0.00}^{0.02}$ & $10.19_{0.03}^{0.03}$ & $1.2_{0.4}^{0.4}$ & $3.4_{0.4}^{0.3}$ \\
GSD-45972 & $1.041_{0.002}^{0.001}$ & $10.94_{0.01}^{0.01}$ & $-11.9_{0.4}^{0.3}$ & $0.00_{0.00}^{0.02}$ & $9.77_{0.03}^{0.03}$ & $2.3_{0.7}^{0.8}$ & $4.5_{1.2}^{2.3}$ \\
GSD-44620 & $1.083_{0.001}^{0.004}$ & $10.72_{0.03}^{0.02}$ & $-11.8_{0.3}^{0.1}$ & $0.05_{0.01}^{0.09}$ & $9.80_{0.03}^{0.03}$ & $0.5_{0.4}^{0.5}$ & $6.9_{1.5}^{1.4}$ \\
GSD-29928 & $1.098_{0.001}^{0.001}$ & $11.67_{0.01}^{0.01}$ & $-12.1_{0.4}^{0.3}$ & $0.35_{0.31}^{0.36}$ & $10.49_{0.03}^{0.03}$ & $1.5_{0.6}^{0.5}$ & $5.1_{1.3}^{1.6}$ \\
GND-30358 & $1.104_{0.005}^{0.005}$ & $10.59_{0.03}^{0.02}$ & $-12.6_{0.7}^{0.4}$ & $0.03_{0.01}^{0.09}$ & $9.77_{0.04}^{0.04}$ & $1.4_{0.5}^{0.3}$ & $2.5_{0.3}^{0.2}$ \\
GND-23857 & $1.121_{0.014}^{0.031}$ & $10.53_{0.03}^{0.02}$ & $-11.6_{0.5}^{0.1}$ & $0.00_{0.00}^{0.03}$ & $9.64_{0.03}^{0.03}$ & $1.7_{0.9}^{0.4}$ & $3.2_{0.8}^{0.4}$ \\
GSD-47691 & $1.127_{0.005}^{0.002}$ & $11.09_{0.02}^{0.03}$ & $-11.5_{0.4}^{0.3}$ & $0.36_{0.30}^{0.49}$ & $9.87_{0.04}^{0.04}$ & $1.7_{0.9}^{0.2}$ & $2.9_{0.8}^{0.1}$ \\
GND-21724 & $1.133_{0.005}^{0.007}$ & $10.89_{0.03}^{0.02}$ & $-11.8_{0.3}^{0.4}$ & $0.25_{0.18}^{0.44}$ & $9.92_{0.04}^{0.04}$ & $1.2_{0.6}^{0.4}$ & $2.9_{0.5}^{0.3}$ \\
GND-37325 & $1.136_{0.009}^{0.009}$ & $10.53_{0.04}^{0.04}$ & $-11.2_{0.5}^{0.3}$ & $0.63_{0.40}^{0.74}$ & $9.82_{0.04}^{0.04}$ & $1.8_{0.8}^{0.3}$ & $2.6_{0.6}^{0.3}$ \\
GND-22027 & $1.141_{0.002}^{0.004}$ & $10.83_{0.03}^{0.03}$ & $-12.0_{0.2}^{0.5}$ & $0.00_{0.00}^{0.17}$ & $9.93_{0.04}^{0.04}$ & $1.1_{0.6}^{0.5}$ & $3.1_{0.6}^{0.4}$ \\
GND-34694 & $1.142_{0.002}^{0.003}$ & $11.15_{0.04}^{0.04}$ & $-11.4_{0.3}^{0.3}$ & $0.32_{0.22}^{0.38}$ & $9.91_{0.06}^{0.06}$ & $1.9_{0.5}^{0.5}$ & $3.3_{0.5}^{0.7}$ \\
GND-38102 & $1.145_{0.009}^{0.008}$ & $10.56_{0.03}^{0.03}$ & $-11.7_{0.4}^{0.4}$ & $0.00_{0.00}^{0.15}$ & $9.91_{0.04}^{0.04}$ & $1.8_{0.9}^{0.5}$ & $2.8_{0.7}^{0.6}$ \\
GND-28451 & $1.148_{0.006}^{0.007}$ & $10.62_{0.02}^{0.02}$ & $-11.7_{0.2}^{0.4}$ & $0.02_{0.01}^{0.04}$ & $9.83_{0.03}^{0.03}$ & $1.5_{0.7}^{0.4}$ & $2.5_{0.4}^{0.4}$ \\
GND-20432 & $1.149_{0.006}^{0.005}$ & $10.94_{0.02}^{0.04}$ & $-11.8_{0.2}^{0.4}$ & $0.19_{0.14}^{0.32}$ & $10.14_{0.03}^{0.03}$ & $1.5_{0.5}^{0.5}$ & $4.7_{1.0}^{1.5}$ \\
GND-17746 & $1.152_{0.009}^{0.008}$ & $11.06_{0.04}^{0.04}$ & $-12.1_{0.3}^{0.5}$ & $0.35_{0.26}^{0.45}$ & $10.24_{0.05}^{0.05}$ & $1.5_{0.6}^{0.4}$ & $5.4_{1.2}^{1.5}$ \\
GSD-39805 & $1.156_{0.012}^{0.017}$ & $10.64_{0.02}^{0.04}$ & $-12.0_{0.3}^{0.4}$ & $0.08_{0.03}^{0.18}$ & $9.70_{0.04}^{0.04}$ & $1.7_{0.5}^{0.3}$ & $5.3_{1.4}^{0.7}$ \\
GSD-40476 & $1.212_{0.001}^{0.003}$ & $10.74_{0.08}^{0.02}$ & $-12.8_{0.7}^{0.6}$ & $0.25_{0.20}^{0.32}$ & $9.95_{0.05}^{0.05}$ & $0.9_{0.7}^{0.5}$ & $2.4_{0.4}^{0.4}$ \\
GSD-37828 & $1.213_{0.003}^{0.002}$ & $10.61_{0.03}^{0.03}$ & $-11.9_{0.4}^{0.3}$ & $0.23_{0.12}^{0.30}$ & $9.85_{0.03}^{0.03}$ & $1.7_{0.6}^{0.3}$ & $3.3_{0.6}^{0.4}$ \\
GSD-40597 & $1.219_{0.003}^{0.001}$ & $11.15_{0.02}^{0.02}$ & $-12.4_{0.7}^{0.4}$ & $0.52_{0.47}^{0.57}$ & $10.21_{0.03}^{0.03}$ & $1.0_{0.3}^{0.1}$ & $2.4_{0.2}^{0.1}$ \\
GND-34419 & $1.221_{0.003}^{0.003}$ & $10.69_{0.03}^{0.03}$ & $-12.2_{0.5}^{0.5}$ & $0.01_{0.01}^{0.04}$ & $9.99_{0.04}^{0.04}$ & $1.1_{0.7}^{0.2}$ & $2.2_{0.4}^{0.1}$ \\
GND-13191 & $1.221_{0.013}^{0.015}$ & $10.68_{0.03}^{0.04}$ & $-11.5_{0.3}^{0.6}$ & $0.57_{0.43}^{0.73}$ & $9.88_{0.04}^{0.04}$ & $1.0_{0.4}^{0.2}$ & $2.3_{0.3}^{0.2}$ \\
GND-14713 & $1.228_{0.003}^{0.004}$ & $10.79_{0.01}^{0.01}$ & $-11.5_{0.3}^{0.1}$ & $0.00_{0.00}^{0.02}$ & $9.93_{0.03}^{0.03}$ & $0.3_{0.3}^{0.2}$ & $3.5_{0.4}^{0.1}$ \\
GND-17070 & $1.238_{0.002}^{0.004}$ & $11.15_{0.03}^{0.02}$ & $-12.0_{0.2}^{0.4}$ & $0.00_{0.00}^{0.02}$ & $10.36_{0.04}^{0.04}$ & $1.5_{0.6}^{0.5}$ & $4.3_{0.9}^{0.6}$ \\
GSD-38785 & $1.241_{0.010}^{0.011}$ & $11.01_{0.04}^{0.03}$ & $-11.0_{0.2}^{0.4}$ & $0.21_{0.16}^{0.30}$ & $9.77_{0.05}^{0.05}$ & $1.6_{0.6}^{0.3}$ & $3.3_{0.6}^{0.5}$ \\
GND-21156 & $1.254_{0.002}^{0.002}$ & $11.30_{0.01}^{0.02}$ & $-11.3_{0.3}^{0.1}$ & $0.33_{0.28}^{0.38}$ & $10.24_{0.03}^{0.03}$ & $1.6_{1.1}^{0.5}$ & $3.4_{1.1}^{0.7}$ \\
GSD-35774 & $1.257_{0.002}^{0.003}$ & $11.21_{0.03}^{0.01}$ & $-12.0_{0.3}^{0.4}$ & $0.02_{0.01}^{0.03}$ & $10.12_{0.03}^{0.03}$ & $1.4_{0.4}^{0.3}$ & $3.6_{0.4}^{0.4}$ \\
GND-37686 & $1.259_{0.003}^{0.001}$ & $11.11_{0.03}^{0.03}$ & $-12.2_{0.2}^{0.7}$ & $0.03_{0.01}^{0.09}$ & $10.23_{0.03}^{0.03}$ & $0.5_{1.0}^{0.3}$ & $3.5_{0.7}^{0.2}$ \\
GSD-40862 & $1.333_{0.004}^{0.005}$ & $11.11_{0.02}^{0.02}$ & $-10.9_{0.3}^{0.1}$ & $0.45_{0.41}^{0.50}$ & $9.90_{0.05}^{0.05}$ & $1.5_{0.2}^{0.1}$ & $3.2_{0.2}^{0.2}$ \\
GSD-46066 & $1.333_{0.003}^{0.003}$ & $11.10_{0.02}^{0.02}$ & $-12.5_{0.3}^{0.1}$ & $0.00_{0.00}^{0.01}$ & $10.20_{0.03}^{0.03}$ & $0.3_{0.2}^{0.2}$ & $6.5_{1.0}^{0.5}$ \\
GSD-39804 & $1.339_{0.002}^{0.004}$ & $11.09_{0.02}^{0.02}$ & $-11.5_{0.1}^{0.3}$ & $0.34_{0.23}^{0.37}$ & $10.35_{0.04}^{0.04}$ & $1.0_{0.4}^{0.2}$ & $3.0_{0.3}^{0.2}$ \\
GSD-45775 & $1.352_{0.007}^{0.005}$ & $11.28_{0.02}^{0.03}$ & $-11.3_{0.1}^{0.1}$ & $0.08_{0.05}^{0.13}$ & $9.74_{0.04}^{0.04}$ & $1.4_{0.4}^{0.2}$ & $5.8_{1.2}^{0.5}$ \\
GND-36530 & $1.362_{0.002}^{0.003}$ & $11.02_{0.01}^{0.01}$ & $-11.3_{0.1}^{0.5}$ & $0.00_{0.00}^{0.01}$ & $10.03_{0.03}^{0.03}$ & $0.7_{0.2}^{0.1}$ & $3.3_{0.3}^{0.1}$ \\
GSD-40623 & $1.414_{0.003}^{0.005}$ & $11.04_{0.01}^{0.02}$ & $-12.1_{0.4}^{0.1}$ & $0.05_{0.01}^{0.08}$ & $10.00_{0.03}^{0.03}$ & $0.3_{0.4}^{0.3}$ & $9.3_{1.8}^{1.2}$ \\
GND-24345 & $1.415_{0.014}^{0.012}$ & $10.57_{0.04}^{0.01}$ & $-11.6_{0.2}^{0.5}$ & $0.00_{0.00}^{0.04}$ & $9.92_{0.03}^{0.03}$ & $0.9_{0.2}^{0.2}$ & $4.9_{0.6}^{0.6}$ \\
GND-16574 & $1.456_{0.005}^{0.005}$ & $10.58_{0.04}^{0.03}$ & $-11.7_{0.5}^{0.2}$ & $0.18_{0.05}^{0.34}$ & $9.89_{0.04}^{0.04}$ & $1.0_{0.3}^{0.3}$ & $3.1_{0.3}^{0.3}$ \\
GND-21427 & $1.472_{0.009}^{0.010}$ & $10.92_{0.02}^{0.02}$ & $-11.4_{0.3}^{0.4}$ & $0.31_{0.24}^{0.42}$ & $10.06_{0.03}^{0.03}$ & $1.1_{0.4}^{0.1}$ & $2.7_{0.4}^{0.1}$ \\
GSD-40223 & $1.599_{0.002}^{0.004}$ & $10.87_{0.02}^{0.05}$ & $-11.3_{0.3}^{0.3}$ & $0.41_{0.32}^{0.46}$ & $10.06_{0.05}^{0.05}$ & $1.2_{0.6}^{0.2}$ & $4.4_{1.0}^{0.5}$ \\
GSD-39649 & $1.603_{0.001}^{0.002}$ & $10.75_{0.03}^{0.02}$ & $-12.1_{0.5}^{0.5}$ & $0.29_{0.21}^{0.39}$ & $10.04_{0.03}^{0.03}$ & $1.3_{0.4}^{0.2}$ & $3.2_{0.4}^{0.3}$ \\
GSD-42487 & $1.605_{0.001}^{0.001}$ & $10.94_{0.02}^{0.03}$ & $-11.6_{0.2}^{0.4}$ & $0.42_{0.31}^{0.49}$ & $10.25_{0.03}^{0.03}$ & $1.1_{0.3}^{0.2}$ & $3.1_{0.4}^{0.2}$ \\
GSD-38843 & $1.611_{0.006}^{0.006}$ & $10.54_{0.04}^{0.04}$ & $-12.1_{0.5}^{0.5}$ & $0.31_{0.19}^{0.40}$ & $9.69_{0.05}^{0.05}$ & $1.4_{0.5}^{0.3}$ & $4.3_{1.0}^{0.8}$ \\
GSD-39012 & $1.612_{0.005}^{0.006}$ & $11.15_{0.02}^{0.02}$ & $-11.7_{0.1}^{0.4}$ & $0.49_{0.44}^{0.56}$ & $10.20_{0.03}^{0.03}$ & $0.3_{0.8}^{0.3}$ & $7.4_{1.4}^{1.2}$ \\
GSD-41520 & $1.614_{0.004}^{0.001}$ & $11.08_{0.03}^{0.02}$ & $-11.6_{0.2}^{0.4}$ & $0.24_{0.17}^{0.30}$ & $10.29_{0.04}^{0.04}$ & $1.1_{0.3}^{0.4}$ & $4.1_{0.3}^{0.8}$ \\
GSD-44042 & $1.616_{0.003}^{0.003}$ & $11.22_{0.04}^{0.03}$ & $-11.1_{0.1}^{0.2}$ & $0.63_{0.57}^{0.70}$ & $10.24_{0.04}^{0.04}$ & $0.8_{0.3}^{0.2}$ & $3.3_{0.4}^{0.3}$ \\
GND-33775 & $1.652_{0.006}^{0.005}$ & $10.59_{0.04}^{0.05}$ & $-11.3_{0.4}^{0.6}$ & $0.25_{0.19}^{0.32}$ & $9.92_{0.04}^{0.04}$ & $1.0_{1.0}^{0.2}$ & $4.0_{1.1}^{0.5}$ \\
GSD-42615 & $1.755_{0.004}^{0.004}$ & $11.03_{0.04}^{0.03}$ & $-12.3_{0.5}^{0.4}$ & $0.00_{0.00}^{0.31}$ & $10.26_{0.04}^{0.04}$ & $1.0_{0.4}^{0.4}$ & $6.3_{1.6}^{2.0}$ \\
GSD-41148 & $1.763_{0.003}^{0.002}$ & $11.22_{0.03}^{0.04}$ & $-11.8_{0.4}^{0.3}$ & $0.33_{0.26}^{0.41}$ & $10.21_{0.04}^{0.04}$ & $0.9_{0.4}^{0.3}$ & $4.1_{0.6}^{0.6}$ \\
GND-33780 & $1.876_{0.016}^{0.016}$ & $11.37_{0.03}^{0.03}$ & $-11.6_{0.6}^{0.6}$ & $0.00_{0.00}^{0.05}$ & $10.19_{0.04}^{0.04}$ & $0.7_{0.8}^{0.2}$ & $4.0_{1.1}^{0.2}$ \\
GND-17735 & $1.876_{0.014}^{0.009}$ & $11.08_{0.03}^{0.02}$ & $-11.3_{0.5}^{0.4}$ & $0.17_{0.11}^{0.26}$ & $10.26_{0.04}^{0.04}$ & $0.9_{0.4}^{0.2}$ & $3.3_{0.5}^{0.3}$ \\
GND-19850 & $1.876_{0.007}^{0.006}$ & $10.81_{0.02}^{0.03}$ & $-12.1_{0.4}^{0.4}$ & $0.00_{0.00}^{0.02}$ & $10.09_{0.03}^{0.03}$ & $1.2_{0.3}^{0.2}$ & $4.5_{0.7}^{0.6}$ \\
GSD-24569 & $1.901_{0.002}^{0.001}$ & $10.83_{0.02}^{0.01}$ & $-11.6_{0.3}^{0.3}$ & $0.24_{0.17}^{0.31}$ & $10.20_{0.02}^{0.02}$ & $0.7_{0.2}^{0.1}$ & $2.9_{0.2}^{0.1}$ \\
GSD-24315 & $1.988_{0.003}^{0.004}$ & $10.54_{0.02}^{0.02}$ & $-11.5_{0.7}^{0.5}$ & $0.41_{0.25}^{0.52}$ & $9.93_{0.03}^{0.03}$ & $0.6_{0.3}^{0.1}$ & $2.9_{0.3}^{0.2}$ \\
GND-14132 & $2.017_{0.052}^{0.015}$ & $11.02_{0.02}^{0.03}$ & $-11.3_{0.3}^{0.3}$ & $0.00_{0.00}^{0.03}$ & $10.18_{0.03}^{0.03}$ & $1.0_{0.3}^{0.2}$ & $4.5_{0.6}^{0.4}$ \\
GSD-43572 & $2.057_{0.039}^{0.034}$ & $11.05_{0.03}^{0.04}$ & $-11.0_{0.2}^{0.5}$ & $0.31_{0.20}^{0.43}$ & $9.88_{0.05}^{0.05}$ & $1.0_{0.3}^{0.2}$ & $5.4_{0.9}^{1.0}$ \\
GND-21738 & $2.092_{0.008}^{0.007}$ & $11.19_{0.06}^{0.04}$ & $-11.9_{0.4}^{0.7}$ & $0.07_{0.04}^{0.15}$ & $10.36_{0.06}^{0.06}$ & $1.0_{0.4}^{0.3}$ & $5.8_{1.2}^{1.2}$ \\
GND-32933 & $2.131_{0.017}^{0.019}$ & $10.71_{0.03}^{0.03}$ & $-10.3_{0.3}^{0.2}$ & $0.36_{0.26}^{0.47}$ & $9.94_{0.04}^{0.04}$ & $1.1_{0.4}^{0.1}$ & $4.6_{0.9}^{0.3}$ \\
GND-17599 & $2.140_{0.002}^{0.002}$ & $10.88_{0.03}^{0.03}$ & $-12.0_{0.4}^{0.7}$ & $0.16_{0.09}^{0.27}$ & $10.30_{0.03}^{0.03}$ & $0.8_{0.3}^{0.3}$ & $4.4_{0.6}^{0.9}$ \\
GSD-44133 & $2.184_{0.006}^{0.005}$ & $10.61_{0.02}^{0.02}$ & $-10.7_{0.3}^{0.5}$ & $0.33_{0.22}^{0.38}$ & $9.62_{0.04}^{0.04}$ & $0.9_{0.4}^{0.1}$ & $3.6_{0.6}^{0.1}$ \\
GND-23018 & $2.302_{0.016}^{0.017}$ & $11.24_{0.04}^{0.03}$ & $-10.8_{0.1}^{0.1}$ & $0.07_{0.05}^{0.09}$ & $10.15_{0.04}^{0.04}$ & $0.7_{0.2}^{0.1}$ & $7.6_{1.4}^{1.0}$ \\
GSD-48464 & $2.349_{0.012}^{0.007}$ & $11.22_{0.02}^{0.02}$ & $-10.7_{0.2}^{0.1}$ & $0.34_{0.30}^{0.40}$ & $10.31_{0.03}^{0.03}$ & $0.8_{0.2}^{0.1}$ & $4.9_{0.5}^{0.2}$ \\
\enddata
\tablecomments{\editone{(1) catalog ID number (matching those in
    \citet{skel14}) and line-matched to those in Table~\ref{table1};
    All other quantities are derived from the model fits to the full
    grism and photometric dataset.   
(2) redshift; (3) stellar  mass; (4) specific star-formation rate
(where the SFR is the time averaged over the previous 100 Myr of the
SFH); (5) dust attenuation $A_V$ value for a Milky Way dust law; (6)
stellar mass surface density within 1~kpc, \Lsig; 
(7) quenching timescale defined as $t_{50}$ - $t_{90}$, the difference
between the time when the galaxy had formed 50\% ($t_{50}$) and 90\%
($t_{90}$) of its stellar mass; (8) formation
redshift (where the galaxy had formed 50\% of its stellar mass); 
Note that we are using a highest density region to estimate our parameter fits, this 
reports the mode and smallest region containing 68$\%$ of the
probability \citep{bail18}. Therefore
if the mode of the probability distribution function is peaked at the
bounds on the parameter, then the uncertainty will also be zero beyond
that bound.  This is the case for some values of $A_V$, for example, where
the mode of the distribution function is $A_V$=0.0 mag (and the lower
68\%-tile uncertainty is likewise 0.0 mag).  }}
\end{deluxetable*}

{~~~~}

\end{document}